\documentclass[twocolumn,tighten,twocolappendix]{aastex631}

\usepackage{newtxtext}
\usepackage[smallerops,cmbraces]{newtxmath}
\usepackage{anyfontsize} 

\makeatletter

\def\frontmatter@title@format{
\Large
\bf\centering
}

\def\frontmatter@authorformat{\normalsize\advance\baselineskip-3pt
\parskip=0pt
\ifmodern
\vskip48pt
\centering
\leftskip=.05in plus 1fil
\rightskip=.05in plus 1 fil
\baselineskip=13pt
\else
\centering
\fi
}%

\def\frontmatter@affiliationfont{\normalfont\footnotesize
\iflongauthor\else
\ifrnaas\else
\rightskip-12pt plus 1fil
\leftskip6pt \parindent-4pt
\fi\fi
}%

\def\abstractname{\normalsize\bfseries Abstract}

\def\sec@upcase#1{\textbf{#1}}
\def\subsec@upcase#1{\relax{#1}}

\def\@sect#1#2#3#4#5#6[#7]#8%
{\ifnum#2=1\setbox0=\hbox{\def\label##1{\gdef\templabel{##1}}#7}\fi
\@tempskipa #5\relax 
 \ifdim \@tempskipa >\z@ \begingroup
     #6\relax 
  \ifnum #2>\c@secnumdepth \def \@svsec {}\else 
    \refstepcounter{#1} \edef \@svsec {\ApjSectionMarkInTitle 
    {\csname the#1\endcsname}}\fi
  \@hangfrom {\hskip #3\relax 
    \ifnum #2=1{\secnum@size {\bfseries\@svsec~}}%
    \else %
        \ifnum #2=2{\subsecnum@size {\itshape\@svsec~}}%
        \else {{\itshape\@svsec~}}\fi%
    \fi }%
  {\interlinepenalty \@M 
   \ifnum #2=1\sec@upcase{#8}%
   \else \subsec@upcase{#8}\fi\par}\endgroup 
  \csname #1mark\endcsname {#7}\addcontentsline{toc}{#1}%
  {\ifnum #2>\c@secnumdepth \else \protect \numberline 
     {\csname the#1\endcsname }\fi #7}%
  \else 
  \ifnum #2>\c@secnumdepth \def \@svsec {}\else 
    \refstepcounter{#1} \edef \@svsec {\ApjSectionMarkInTitle 
    {\csname the#1\endcsname}}\fi
    \def \@svsechd {#6\hskip #3%
    \ifnum #2=1{\secnum@size{\bfseries\@svsec~}}\else{\subsecnum@size{\bfseries\@svsec~}}\fi%
    \ifnum #2=1\sec@upcase{#8}\else\subsec@upcase{#8}\fi%
    \ifnum #2=4\hskip 0.4ex{\rm ---}\fi%
     \csname #1mark\endcsname {#7}\addcontentsline{toc}{#1}%
      {\ifnum #2>\c@secnumdepth \else \protect \numberline {\csname
      the#1\endcsname }\fi #7}}\fi \@xsect {#5} \penalty \ApjSectionpenalty
\protected@edef\@currentlabel{\csname p@#1\endcsname\csname
the#1\endcsname} 
\expandafter\ifx\csname templabel\endcsname\relax
\let\go\relax
\else
\def\go{\label{\templabel}}\fi\go
\let\templabel\relax
}

\def\acknowledgments{%
\global\setbox\ackbox=\vbox\bgroup
\vskip 5.8mm plus 1mm minus 1mm
\vskip1sp
\noindent\ignorespaces}

\def\endacknowledgments{
\egroup
\ifanonymous
\vskip 5.8mm plus 1mm minus 1mm
\vskip1sp
\centerline{(Acknowledgments anonymized for review)}
\else
\vbox{\unvbox\ackbox}
\fi\vskip6pt}

\def\appendix{\global\appendixontrue
\if@two@col  
\onecolumngrid
\noindent\mbox{}\vrule height 24pt width0pt\hfill{\apjsecfont\bfseries Appendix}\hfill\mbox{}\par
\vskip18pt
   \if@two@col@app\global\@two@coltrue\twocolumngrid \fi
\else
\noindent\mbox{}\vrule height 24pt width0pt\hfill{\apjsecfont\bfseries
Appendix}\hfill\mbox{}\par 
\vskip18pt
    \if@two@col@app\global\@two@coltrue\twocolumngrid 
         \fi\fi
        \nopagebreak\medskip\@nobreaktrue\def\ApjSectionpenalty{\@M}
        \@firstsectionfalse
          \setcounter{section}{0}
          \setcounter{subsection}{0}
          \setcounter{equation}{0}
\ifletteredappendix
          \def\thesection{\Alph{section}}
          \def\theequation{\hbox{\Alph{section}\arabic{equation}}}
\fi
\if@number@appendix@floats
          \def\thesection{\Alph{section}}
          \def\theequation{\hbox{\Alph{section}\arabic{equation}}}
          \def\section{\@startsection {section}{1}{\z@} 
            {10pt}{5pt}{\centering\scshape\apjsecfont}}
\else
          \def\ApjSectionMarkInTitle{\AppendixApjSectionMarkInTitle}
\fi
\ifappletter
\let\savesection\section
\def\section{\resetapptablenumbers\savesection}
\fi
}

\renewenvironment{thebibliography}[1]{\global\bibtrue
\onecolumngrid
\vspace{20pt}
\goodbreak
    \hbox to\textwidth{\hss\normalsize\bfseries References\hss} 
\vspace{6pt}\parskip=0pt
\twocolumngrid 
\ifnumlines
\ifonecolstyle
\ifpreprinttwo\else
\advance\linenumbersep-12pt\fi\fi\fi
\par
 \raggedright
\small
\ifmodern\else
 \vspace{10pt plus 3pt}\fi
\par
\topsep=0pt
 \list{}%
   {
     \parindent=0pt \parskip=1pt plus 1pt \parsep=0pt 
     \bibindent=0pt                          %
\ifmodern\vskip-12pt
\baselineskip=13pt plus 1pt
\else
\ifdoublespace
\baselineskip=20pt
\else
\baselineskip=13pt plus 1pt \fi\fi \interlinepenalty \@M  
     \frenchspacing    
     \hyphenpenalty=10000
     \itemindent=-1.0em                      %
     \itemsep=0pt                            %
     \listparindent=0pt                      %
     \settowidth\labelwidth{0pt} %
     \labelsep=0pt                           %
     \leftmargin=1.0em
     \advance\leftmargin\labelsep
      \let\p@enumiv\@empty
      }%
    \sloppy\clubpenalty10000\widowpenalty10000%
    \sfcode`\.\@m\relax
}

\makeatother

\usepackage{etoolbox}

\usepackage{bm} 
\usepackage{textgreek} 

\usepackage[capitalize,noabbrev]{cleveref} 
\creflabelformat{equation}{#2#1#3} 
\crefname{enumi}{item}{items} 

\makeatletter
\patchcmd\H@refstepcounter{\protected@edef}{\protected@xdef}{}{}
\makeatother

\let\tablenum\relax
\usepackage{siunitx} 
\DeclareSIUnit[number-unit-product = ]\percent{\char`\%} 

\usepackage{placeins}
\usepackage{csquotes}

\definecolor{blackberry}{HTML}{8D1D75}

\newcommand*{\m}[1]{M\,#1}

\DeclareSIUnit\parsec{pc}
\DeclareSIUnit\dex{dex}
\DeclareSIUnit\h{\mathnormal{h}}
\DeclareSIUnit\year{yr}
\DeclareSIUnit\years{yrs}
\DeclareSIUnit\arcsec{arcsec}
\DeclareSIUnit\arcmin{arcmin}
\DeclareSIUnit\Msun{M_\odot}
\DeclareSIUnit\Rsun{R_\odot}
\DeclareSIUnit\Lsun{L_\odot}
\DeclareSIUnit\Rvir{\mathnormal{R}_\mathrm{vir}}
\DeclareSIUnit\Rhalf{\mathnormal{R}_{1/2}}
\DeclareSIUnit\erg{erg}
\DeclareSIUnit\angstrom{\text{Å}}

\newcommand*{\Rc}{\ensuremath{\mathrm{R}_\mathrm{200,c}}}

\newcommand*{\Msun}{\ensuremath{\mathrm{M}_\odot}} 
\newcommand*{\Rsun}{\ensuremath{\mathrm{R}_\odot}} 
\newcommand*{\Lsun}{\ensuremath{\mathrm{L}_\odot}} 
\newcommand*{\Rvir}{\ensuremath{R_\mathrm{vir}}} 
\newcommand*{\Rhalf}{\ensuremath{R_{1/2}}} 

\newcommand*{\eshort}[2]{\ensuremath{#1 \times 10^{#2}}}

\let\oldbibliography\thebibliography
\renewcommand{\thebibliography}[1]{%
  \oldbibliography{#1}%
  \setlength{\itemsep}{1.95pt}%
  \setlength{\baselineskip}{10pt}
  \setlength{\lineskiplimit}{-\maxdimen}
}

\definecolor{lightblue}{rgb}{0.1,0.5,0.89}

\received{-}
\revised{-}
\accepted{-}

\shorttitle{Dwarf Galaxies in Phase-Space}
\shortauthors{Ivleva et al.}

\begin{document}

\title{Phase-Space Diagnostics for Dwarf Galaxies in Cluster Environments}

\correspondingauthor{Anna Ivleva}
\email{ivleva@usm.lmu.de}

\author[0009-0004-8086-8922]{Anna Ivleva}
\affiliation{Universitäts-Sternwarte, Fakultät für Physik, Ludwig-Maximilians-Universität München, Scheinerstr. 1, 81679 München, Germany}

\author[0009-0008-9260-7278]{Rhea-Silvia Remus}
\affiliation{Universitäts-Sternwarte, Fakultät für Physik, Ludwig-Maximilians-Universität München, Scheinerstr. 1, 81679 München, Germany}

\author[0000-0003-1750-286X]{Klaus Dolag}
\affiliation{Universitäts-Sternwarte, Fakultät für Physik, Ludwig-Maximilians-Universität München, Scheinerstr. 1, 81679 München, Germany}
\affiliation{Max-Planck-Institut für Astrophysik, Karl-Scharzschild-Str. 1, 85748 Garching, Germany}

\author[0000-0002-7972-9675]{Lucas M. Valenzuela}
\affiliation{Universitäts-Sternwarte, Fakultät für Physik, Ludwig-Maximilians-Universität München, Scheinerstr. 1, 81679 München, Germany}

\author[0000-0002-2936-7805]{Jonah S. Gannon}
\affiliation{Department of Astronomy \& Astrophysics, University of Toronto, 50 St. George Street, Toronto, ON M5S 3H4, Canada}
\affiliation{Canadian Institute for Theoretical Astrophysics, University of Toronto, 60 St. George Street, Toronto, ON M5S 3H8, Canada}
\affiliation{Dragonfly Focused Research Organization, 150 Washington Avenue, Suite 201, Santa Fe, NM 87501, USA}
\affiliation{Centre for Astrophysics \& Supercomputing, Swinburne University, Hawthorn, VIC 3122, Australia}

\author[0000-0001-5590-5518]{Duncan A. Forbes}
\affiliation{Centre for Astrophysics \& Supercomputing, Swinburne University, Hawthorn, VIC 3122, Australia}

\begin{abstract}

Ongoing effort is devoted to observing spectroscopic samples of dwarf galaxies in clusters, allowing the analysis of their distribution and associated trends in projected phase-space (PPS), i.e. line-of-sight velocity vs. projected clustercentric distance. By utilizing the resolved baryonic halos inside the galaxy clusters of a cosmological simulation from the \textsc{Magneticum} suite, we complement on prior studies with dedicated focus on the dwarf galaxy population ($M_\ast<10^9\,\Msun$) and correlations between infall time and location in PPS. The inferred trend recovers the radial correlation reported by prior works, but we find a significant fraction ($\geq30\%$) of recently accreted galaxies at locations that were previously predicted to be dominated by ancient infallers. Splitting the diagram with an infall time threshold of \SI{3}{\giga\year}, we develop a detailed infall time template in PPS. We provide our data to allow observers to statistically infer the time of infall of their sample when placing them on the PPS. Additionally, we review the trajectories in PPS of different orbits and their dependence on the observer's orientation. Compared to massive galaxies, we find a much broader radial distribution for dwarfs in 3D PS. Utilizing a set of high-resolution idealized simulations, we predict strongly altered orbits for dark matter-deficient galaxies.
\end{abstract}


\section{Introduction}
\label{sec:introduction}

Studying galaxy evolution requires building representative samples of dwarf galaxies, since they are the dominant population in the Universe and constitute the building blocks of more massive galaxies according to the standard hierarchical halo formation paradigm. Due to their overabundance of hosted objects, galaxy clusters naturally arise as the primary target for such assignments. As photometry alone is not sufficient to break degeneracies between various key properties, such as stellar mass, star formation history, metallicity or dust content, spectroscopic surveys become indispensable. The low surface brightness poses an additional challenge here compared to massive galaxies -- driven by the increase in necessary exposure time per target, dwarf galaxy surveys often need to focus on limited areas in the cluster in order to reveal its low surface brightness component. A variety of clusters have now been examined in such a way, e.g. the Coma \citep[][]{Smith:2009,Alabi:2018}, Virgo \citep[][]{Toloba:2009}, Fornax \citep[][]{Iodice:2016,Zabel:2021}, Perseus \citep[][]{Tang:2026} and Hydra cluster \citep[][]{Christlein:2003,Iodice:2023}.

However, the extent and relative importance of internal vs. external influences changes strongly depending on how close a galaxy is to the cluster center. The local environment can induce extensive transformation processes driven by tidal interaction between the galaxy and cluster potential \citep[e.g.,][]{Mastropietro:2005,Aguerri:2009} or in-between galaxies via \enquote{harassment} \citep[][]{Moore:1998}, gas stripping through ram pressure \citep[e.g.,][]{GunnGott:1972,Quilis:2000,Ruszkowski:2014}, and ceased gas accretion called \enquote{strangulation} or \enquote{starvation} \citep[e.g.,][]{Balogh:2000,Peng:2015,vandeVoort:2017}. These conditions impact not only a galaxy's morphology and kinematics \citep[e.g.][]{Lisker:2006,Oxland:2024,Geha:2003,Bagge:2025}, but also lead to strong variations in stellar populations through modified star formation histories \citep[e.g.,][]{Haines:2006,Salvador:2021,Ferremateu:2023}. Hence, it becomes necessary to map the dwarf population with full coverage out to the cluster boundaries of at least its virial radius, where such influences can start to occur, while also taking into account that most clusters are not fully relaxed and therefore not isotropic due to substructures. A few clusters were already targeted in such a full survey with varying depth, in particular the Virgo \citep[][]{Kim:2014}, Perseus \citep[][]{Aguerri:2020}, Hercules \citep[][]{Agulli:2017} and Abell~85 cluster \citep[][]{Agulli:2016}.

Due to their lower gravitational binding energy, dwarf galaxies are expected to be much more susceptible to their environment than massive galaxies and might transform on shorter timescales. This could cause the large number of different categories that dwarf galaxies are being classified in ongoing studies. At fixed stellar mass, dwarf galaxies display a vast diversity of appearances (aside from irregular) and can range from bright spheroidals and ellipticals to conversely low-surface brightnesses. In particular the latter class has received increasing attention in the past decade through the discovery of \enquote{ultra-diffuse galaxies} \citep[UDGs, ][]{vanDokkum:2015}, for which feasible formation pathways are still strongly debated and evidence for diverse origins is accumulating \citep[][]{Alabi:2018,Ferremateu:2018,Gannon:2022,Gannon:2026,Forbes:2023,Tang:2026}

A further complication is that the mass budget of dwarf galaxies is not always necessarily dominated by dark matter. High baryon fractions in a small subgroup of dwarf galaxies has been now reported across different morphological types of dwarfs such as spheroidals \citep[][]{Hammer:2020} and in UDGs \citep[e.g.,][]{vanDokkum:2018,ManceraPina:2019,ForbesGannon2024,Buzzo:2025}, which can also appear in cluster environments. While having similar total masses as regular galaxies, these objects are much more susceptible to destabilizing influences on the gas content through ram pressure, leading to significantly different evolutionary pathways for such dwarfs \citep[][]{Ivleva:2026}.

By now, projected phase-space (PPS) diagrams -- i.e. plotting the observed clustercentric distance of a galaxy vs. its line-of-sight velocity with respect to the cluster -- have proven to be a convenient tool for analyzing galaxy populations in overdense environments, since it encodes the dynamical state of the orbiting halo. Though one cannot infer the full 3D information observationally, several trends have been found to persist throughout the plane despite projection effects, such as stellar mass \citep[][]{Mahajan:2011}, star formation behavior and ram pressure stripping \citep[][]{HernandezFernanedez:2014,Muzzin:2014,Oman:2016,Jaffe:2016,lotz:2019} as well as tidal stripping timescales \citep[e.g.,][]{Smith:2016,Rhee:2017}.

Since a cluster environment has such significant influence on a galaxy, studying galaxy evolution requires quantifying the epoch when it was accreted onto a cluster. For this reason, several dedicated studies have analyzed correlations between infall time into the cluster and location in PPS based on simulated samples, which could help observers to estimate the plausible extent of environmental influences that their galaxy sample might have been exposed to \citep[][]{Mahajan:2011, Oman:2013,Rhee:2017,Pasquali:2019}. However, these studies relied on the dark matter component rather than the resolved baryon halos in order to be independent of assumptions in subgrid models, which are necessary in baryonic simulations. The expected baryon component of the halo and hence the stellar mass limit for galaxies to which these predictions are applicable hence need to be inferred from halo abundance matching models. While these models are now relatively well constrained for massive galaxies, they are not converged at dwarf galaxy mass regimes, where stellar occupation fractions can differ by orders of magnitude at dwarf halo masses \citep[][]{Wechsler:2018}. In addition, probable infall times for particular regions in PPS were assigned by comparing the locations of the maxima of the galaxy distribution in PPS, after dividing the sample into cohorts of different ages. However, when one evaluates the mean based on the whole population hosted in a particular region in PS, the answer could be different driven by local mixing of multiple populations with different orbits and \mbox{infall times}.

In our work, we complement these studies by utilizing the high resolution {\it Box4/uhr} of the \textsc{Magneticum} cosmological simulations \citep[][]{2015ApJ...812...29T,Dolag:2025}, which allows analyzing the infall time distribution for the resolved dwarf galaxy population. Section \ref{sec:sims} summarizes the simulated dwarf galaxy samples used in this work. The overall distribution of dwarfs inside galaxy clusters at redshift $z\approx0$ is presented in Section \ref{sec:cosmo_distrib}. In Section \ref{sec:infalltimeanalysis} we analyze the infall time of dwarf galaxies across PPS and develop an infall time diagnostic tool. We compare and discuss our results in relation to previous studies in Section \ref{sec:discussion_infalltimes}. In Section \ref{sec:dmfreedwarfs} we move towards peculiar dynamical properties exhibited by baryon-dominated populations drawn from high-resolution idealized simulations \citep[][]{Ivleva:2024} and review probable orbital types associated to different regions in PPS. Finally, we summarize our findings in Section \ref{sec:sumcon}.

\section{Simulated dwarf galaxy samples}
\label{sec:sims}

\subsection{Cosmological sample from \textsc{Magneticum}} \label{subsec:magneticum}

Since our goal is to study the dwarf galaxy population inside clusters, a cosmological simulation with very high baryon resolution is necessary. Simultaneously, it needs also a sufficiently large volume (i.e. box length), which determines the size of the largest collapsing node and hence the upper limit of resolved group and cluster masses. {\it Box4/uhr} of the \textsc{Magneticum} suite meets these requirements by a high stellar mass resolution of $M_\ast=\eshort{2}{6}\,\Msun$ and a box length of $68\,\rm Mpc$, yielding three clusters with virial masses of $M_{\rm 200,c}=(1.9, 1.4, 1.3)\times 10^{14}\,\Msun$. We refer to the \textsc{Magneticum} overview paper \citep[][]{Dolag:2025} for an extensive description of the simulation, and outline our selection criteria for the sample examined in this work in the following.

From the three most massive clusters we include all dwarf galaxies with stellar masses between \mbox{$\eshort{2}{8}\,\Msun \leq M_\ast < 10^9\, \Msun$} at redshift $z\approx 0$, applying a resolution limit of about 100  stellar particles. Using these criteria, we arrive at a sample of 639 distinct dwarf galaxies inside the three different clusters. The properties of each halo are based on the structure finding friends-of-friends algorithm \textsc{SubFind} \citep[][]{Springel:2001,Dolag:2009}.

The infall time for each galaxy is defined as the time that has passed since crossing the virial radius $R_{\rm vir}$ for the first time. Note that for the analyzed clusters we have $R_{\rm vir}\approx 1.3R_{\rm 200, c}$. Observational studies of galaxies in clusters generally compare to $R_{\rm 200, c}$ and call this the cluster's virial radius. For continuity purposes we adapt the same nomenclature, and refer to $R_{\rm 200, c}$ as the virial radius throughout this work.

\subsection{Baryon-dominated tidal dwarf galaxies from idealized simulations}\label{subsec:tdgsampleintro}

In addition to the cosmological sample, we investigate the behavior of dark matter-deficient galaxies. For this purpose we utilize high resolution hydrodynamical simulations of tidal dwarf galaxies (TDGs) that were stripped from their parent merger and subsequently populate the cluster \citep[][]{Ivleva:2024,Ivleva:2026}. These simulations naturally ensure an appropriate sample of these peculiar objects formed under realistic conditions and grant fine time resolution allowing for tracing their evolution in PS. The setup and inferred properties of these simulations were presented in detail by \citet[][c.f. setups C25+C45]{Ivleva:2024}, hence we only shortly outline the key features in the following.

The two simulations both start with almost identical setups, modeling the same galaxy merger, which occurs shortly before it crosses the cluster boundary of the virial radius. The only difference is the initial angular momentum of the parent merger, which is on a circular vs. elliptical orbit in the two simulations, respectively. This difference naturally reflects in the orbit of the detached tidal dwarf galaxies and thus allows for a broader dynamical analysis. The baryon resolution for these simulations is $\eshort{6.6}{5}\,\Msun$, resolving dwarf galaxies with total masses between $\eshort{5}{7}\,\Msun \leq M_{\rm halo} \leq 10^9 \,\Msun$, while the cluster has a virial mass of $M_{\rm 200,c}=10^{14}\,\Msun$. The simulations are hydrodynamic and radiative, hence resolving the interaction between the gas component of the galaxy and the ICM, which triggers high star formation rates \citep[with respect to the dwarf main sequence, c.f.][]{Ivleva:2026} and causes the dwarfs to transform from gas-dominated objects with diffuse stellar light to compact ellipticals on timescales of about \SI{4}{\giga\year}.

\section{Distribution of dwarf galaxies in projected phase-space}\label{sec:cosmo_distrib}

\begin{figure}[ht!]
    \centerline{\includegraphics[width=0.5\textwidth, trim={0 0 0 0}]{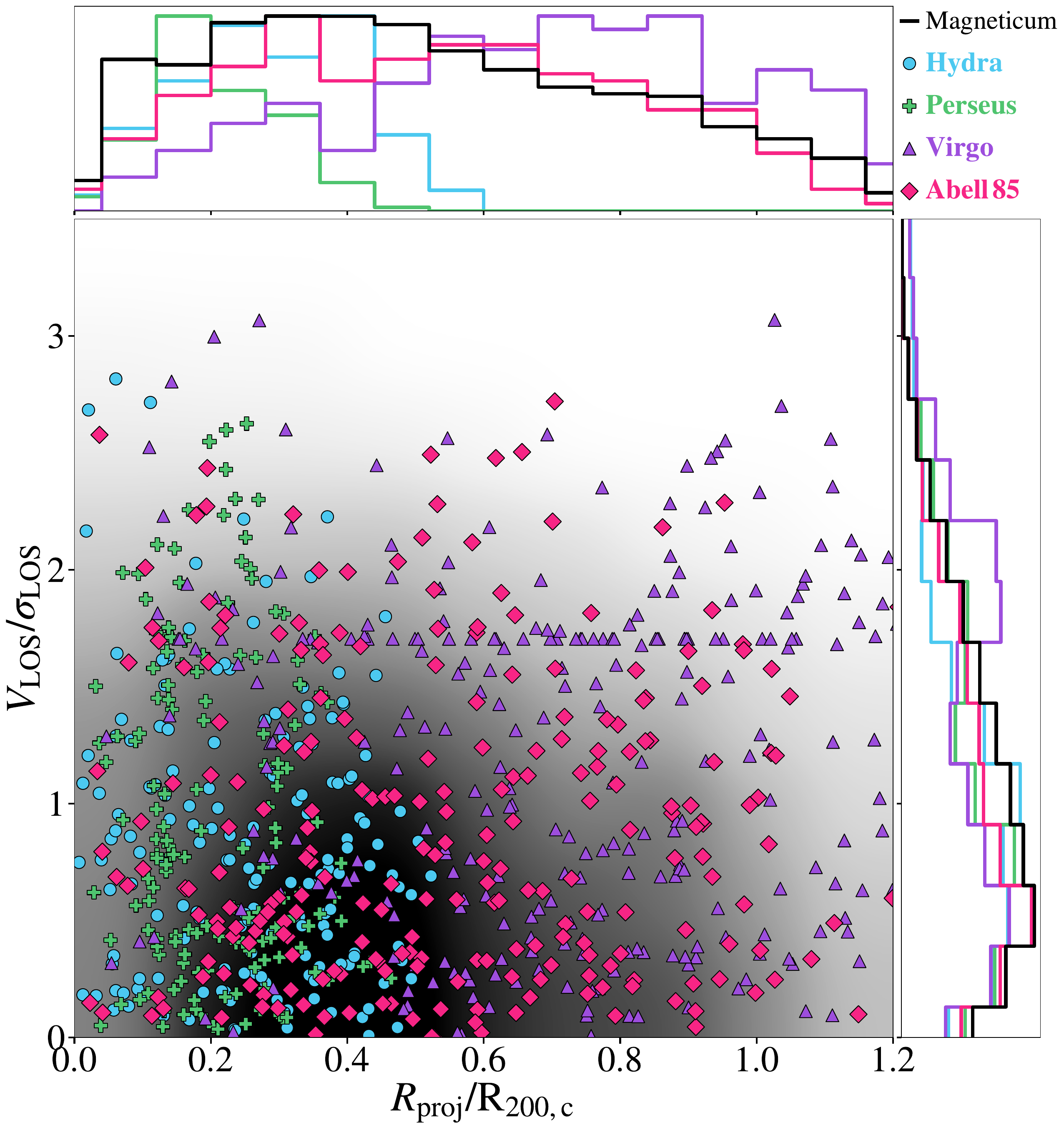}}
    \caption{Number density of projected phase-space distribution of \textsc{Magneticum} dwarf galaxies (grayscale background) compared to observed dwarf samples. The markers represent: Hydra \citep[blue circles,][]{Christlein:2003}, Perseus \citep[green pluses,][]{Tang:2026}, Virgo \citep[violet triangles,][]{Kim:2014}, and A85 \citep[pink diamonds,][]{Agulli:2016}. The adjacent histograms show the respective distributions of each sample.}
    \label{fig:box4diagnostics}
\end{figure} 

We review the distribution of dwarf galaxies predicted by \textsc{Magneticum} in projected phase-space (PPS) and compare to available observations in the following. In order to properly study projection effects, a sufficiently large number $N$ of different line of sights needs to be applied on the simulated sample, where each projection has the same probability to represent an observer's position. This is equivalent to finding a tessellation of the surface of a sphere into $N$ parts that all have the same surface area. This is achieved by the \textsc{HEALPix} pixelation algorithm, which returns $N=12n^2$ such grid cells, where $n$ needs to be an integer of power of 2. The statistical sample of line of sights can thus be constructed by shifting the observer to the position of each cell center, looking towards the midpoint of the sphere, where we place the center of the galaxy cluster defined by its brightest cluster galaxy (BCG). For all our upcoming analysis we choose $n=8$, which results in 768 different  random projections applied to each dwarf galaxy per galaxy cluster.

\cref{fig:box4diagnostics} presents the number density of dwarf galaxies from the cosmological simulation in PPS as gray-scale background. The projected distance $R_{\rm proj}$ and line of sight velocity $V_{\rm LOS}$ with respect to the galaxy cluster center are normalized by the virial radius $\rm R_{\rm 200,c}$ and one-dimensional equivalent of the virial velocity dispersion $\sigma_{\rm LOS} = \sigma_{\rm 200,c} / \sqrt{3} = \sqrt{(G M_{\rm 200,c})/(5R_{\rm 200,c})}$, where $G, M_{\rm 200,c}$ and $R_{\rm 200,c}$ are the gravitational constant and virial mass and radius, respectively. For comparison, we overplot the observed dwarf galaxies from surveys in four different clusters: Hydra \citep[blue circles,][]{Christlein:2003}, Perseus \citep[green pluses,][]{Tang:2026}, Virgo \citep[violet triangles,][]{Kim:2014}, and A85 \citep[pink diamonds,][]{Agulli:2016}. The histograms at the upper and right edge of the panel show the distribution of dwarf galaxies in projected distance and line of sight velocity respectively, where the line color coincides with the markers of the observed samples. The black line denotes the dwarf galaxy distribution from \textsc{Magneticum}. For visibility purposes we normalize each distribution by its maximum, showcasing the qualitative differences between each sample.

For the \textsc{Magneticum} galaxies, the maximum of the distribution function lies at radial distances of about $0.4\,R_{\rm 200,c}$, with a monotonically decreasing tail towards larger radii. While the relative distribution of observed dwarf galaxies in clusters coincides well with the simulated sample in velocity space\footnote{except for bins around $V_{\rm LOS}/\sigma_{\rm LOS}=1.7$, where it seems that an artifact in the observations of the Virgo cluster led to oversampling}, major differences arise with respect to their radial distance to the cluster center. However, there are several biases that need to be noted in this regard. Currently, a systematic caveat of dwarf galaxy surveys in clusters is the limited coverage. In contrast to massive galaxies, dwarfs and -- in particular -- low-surface brightness galaxies have only recently become detectable with currently operating optical telescope facilities. Therefore, most spectroscopic catalogs of dwarf galaxies still cover relatively small fractions of the total cluster volume, focusing only on the brightest, innermost part around the cluster center. As such, the surveys of the Hydra and Perseus cluster shown in \cref{fig:box4diagnostics} reach \enquote{only} until \mbox{$\sim$$0.5\Rc$}, missing a large portion of objects that are predicted to reside at intermediate cluster distances. Meanwhile, though both the Virgo cluster and A85 have coverage until the virial radius, they display qualitatively different distributions in radial direction. A85 agrees relatively well with the simulated distribution predicted by the \textsc{Magneticum} clusters, displaying its maximum around $R\sim0.3\Rc$. The Virgo cluster, on the other hand, exhibits the most dwarf galaxies at much larger radial distances around $0.8\Rc$, having fewer objects in the inner part of the cluster than in the simulated sample.

Aside from possibly insufficient survey completeness, another reason for that difference might be the dynamical state of the cluster. While the Virgo cluster is comparably relaxed, A85 shows clear signs of disturbance \citep{Aguerri:2018} and is thus more similar to the simulated clusters. The underlying reason here is that the clusters in {\it Box4} of \textsc{Magneticum} are generally active since they constitute the largest nodes in the cosmological box. Seeing this trend we stress that further spectroscopic surveys of dwarfs in more galaxy clusters with coverage until the virial radius are necessary for a meaningful analysis of their properties and trends within PPS. Without the population at intermediate cluster radii $>0.5\Rc$ it is not possible to verify tentative trends of varying dwarf features with different cluster properties, since a large portion of the total population hosted by the cluster is missing otherwise.

\section{Infall time diagnostics in projected phase-space}\label{sec:infalltimeanalysis}

This Section is entirely focused on analyzing infall time trends associated with different regions across PPS. We define the infall time $t_{\rm infall}$ as the period that has elapsed since the object has crossed the virial radius $R_{\rm vir}(\approx 1.3\, \rm R_{\rm 200,c})$ of the galaxy cluster. Naturally, we can apply our analysis only to galaxies which have formed ex situ and were subsequently accreted into the cluster. Thus we exclude all objects which formed in situ, which corresponds to about $20\%$ compared to the total population inside each analyzed cluster. Next we apply a Voronoi tesselation to the plane, as this method provides a grid where each cell hosts the same amount of galaxies. This allows to draw statistically meaningful comparisons between populations hosted by different regions across PPS. We choose as the binning criterion for each cell to contain \mbox{$\sim$1000} galaxies. While cross-checking different binning thresholds, we found this value to provide both statistically robust samples in each cell, while also supplying a relatively fine grid in PPS, allowing to study detailed trends with varying PPS location.

\subsection{Infall time for a given region in PPS}
\label{subsec:spread}

\begin{figure}[t!]
    \centerline{\includegraphics[width=0.5\textwidth, trim={0 0 0 0}]{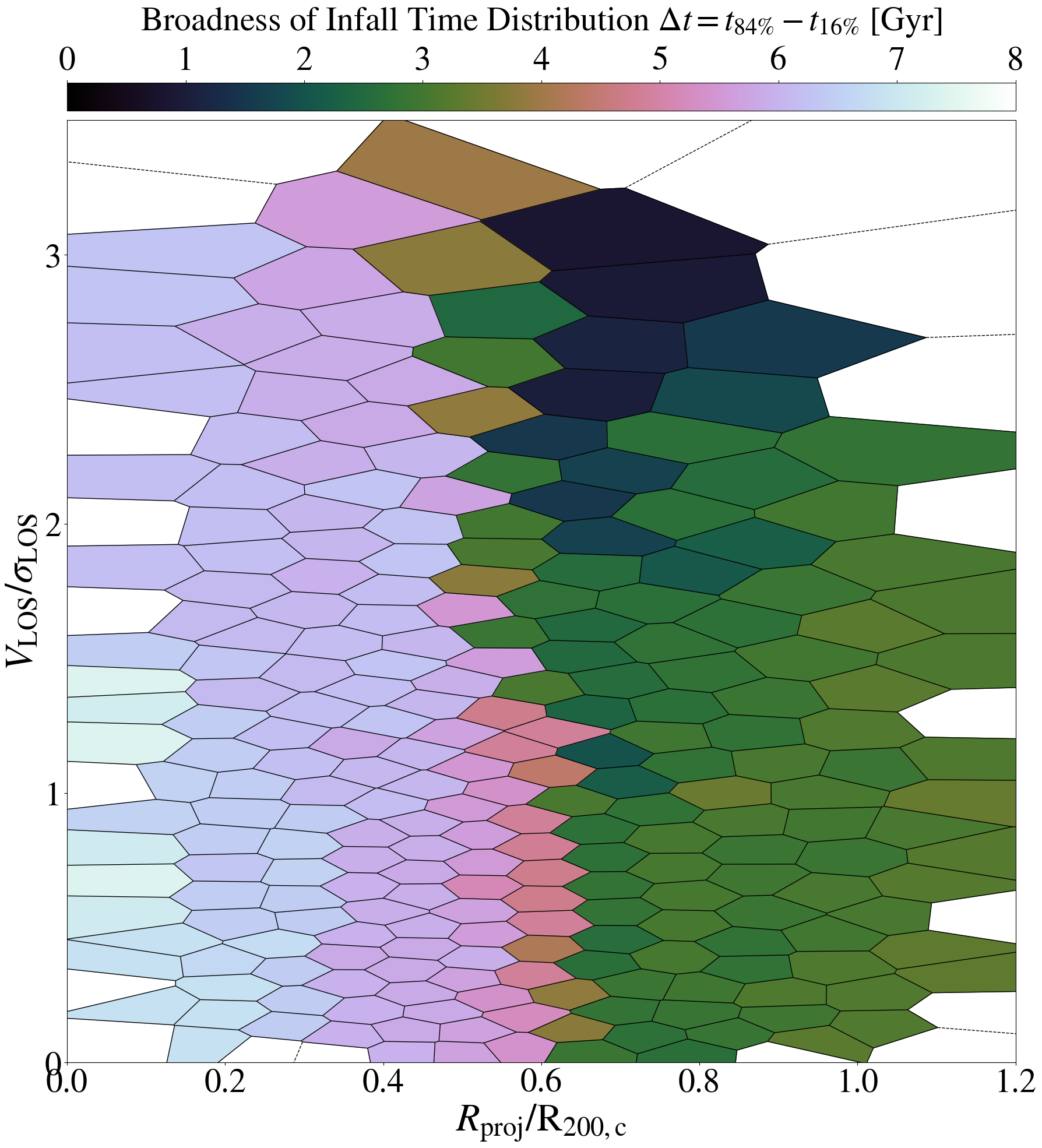}}
    \caption{Broadness of the local distribution of infall times across PPS. The cell color denotes the difference between the 84th and 16th percentile of the galaxies' infall times in each cell. While the outer regions of the cluster have a relatively well defined mean due to small width, the inner parts ($R_{\rm proj}<0.5\Rc$) display a very large spread of infall times associated with these regions.}
    \label{fig:broadness_unimodal}
\end{figure}

Assigning a fixed age to a given region of PPS is only valid if the population within is well described by a unimodal distribution. However, orbits reflect the whole accreation history of galaxy clusters, containing galaxies with multiple pericenter passages and a variety of orbit types. As such, multimodal infall time distributions are expected across PPS, which we investigate in the following.

Following the method described in the beginning of this Section, we constructed a grid for the distribution of the infall times of our dwarfs in PPS. We measure the broadness of the distribution as the difference between the 84th and 16th percentile of the local infall time distribution, which can be interpreted as the extent of the $1\sigma$ scatter around the median. This extent of the distribution is shown as the cell color in \cref{fig:broadness_unimodal}. It immediately becomes apparent that there is extensive variation possible across PPS. While in the outskirts of the cluster the local distribution of infall times is quite confined with variations of less than \SI{2}{\giga\year}, the situation rapidly changes for decreasing clustercentric projected distances. Below $R_{\rm proj}<0.5\Rc$, the local distribution becomes very broad, displaying variations between $5-8\,\rm Gyr$. This indicates a surprisingly large fraction of galaxies with recent infall times at these locations. However, with this information alone it is not possible to discern yet, whether the local infall time distribution is simply flat -- which would not permit to define any meaningful expectancy value for the local infall time in PPS -- or whether the broad distribution is driven by multiple epochs of varying infall time.

\subsection{Fraction of dwarf galaxies with recent infall}
\label{subsec:fraction}

\begin{figure}[t!]
    \centerline{\includegraphics[width=0.5\textwidth, trim={0 0 0 0}]{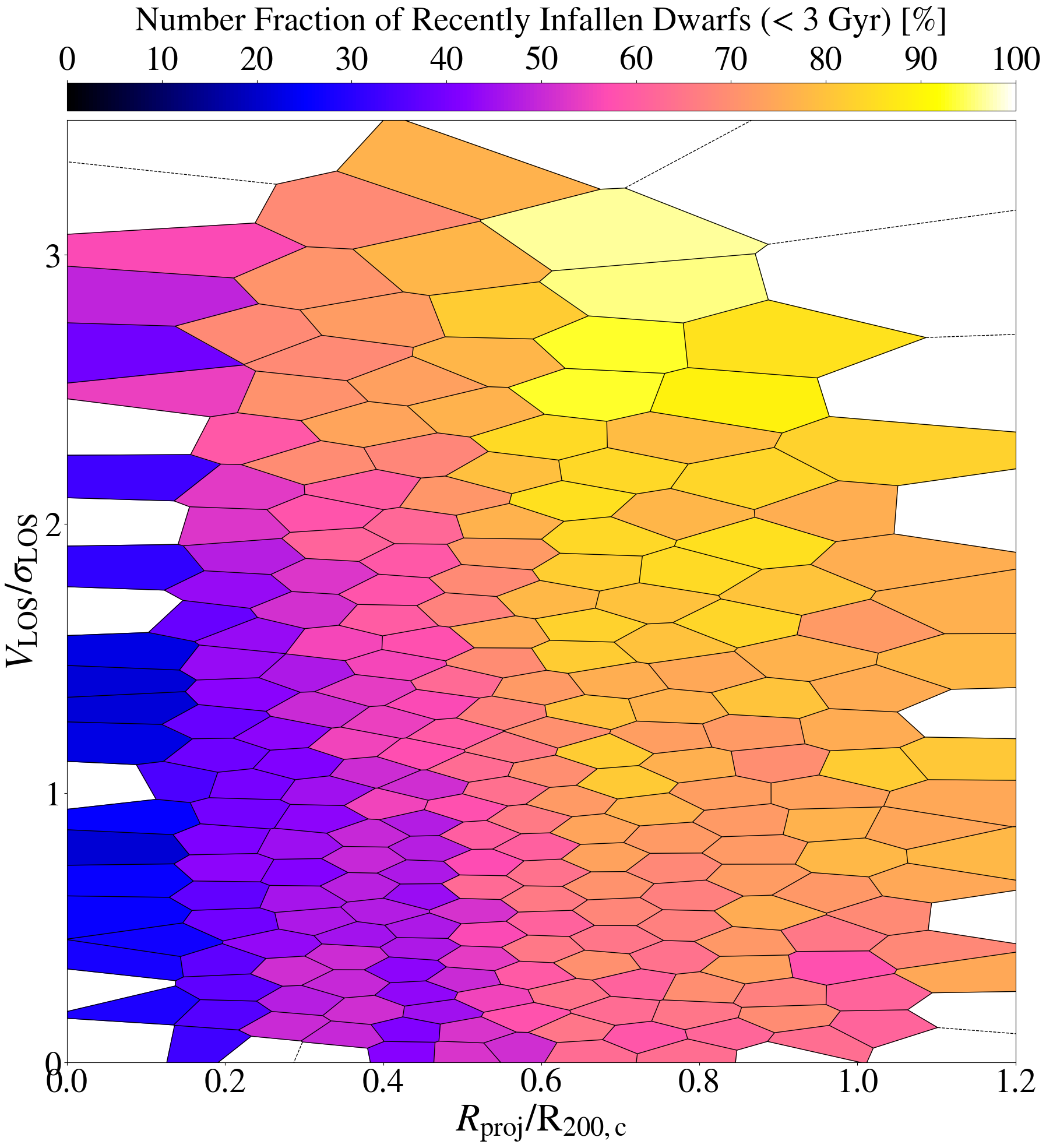}}
    \caption{Fraction of dwarf galaxies with recent infall ($<3\,\rm Gyr$). While the cluster outskirts are almost entirely dominated by recently infallen galaxies, the cluster center does not only host ancient dwarfs, but also at least $30\%$ of recently infallen galaxies.}
    \label{fig:fraction_in_group}
\end{figure}

As a first step, we can split the dwarf galaxy population into a recently vs. early infallen population, by comparing the fraction of galaxies in each cell with infall times smaller than \SI{3}{\giga\year} to the total population in \cref{fig:fraction_in_group}. Cells with large fractions of recently infallen dwarf galaxies as indicated by the colorbar hence show the relative significance of this population at the given location within PPS.

As expected, we find here a clear correlation of more recently infallen dwarfs with increasing clustercentric distance. This is particularly pronounced in the upper right of PPS, since this region can be reached only by dwarfs on first infall, which have not yet experienced strong dynamical friction removing angular momentum. Note, however, that this young (recent) population never drops below $\sim$$30\%$ even in the innermost regions of the cluster. Hence, a significant recent population is expected in addition to the early infallen galaxies throughout the whole galaxy cluster.

Given the broad distribution of infall times as shown in \cref{fig:broadness_unimodal}, it is evident that assigning an unique expected infall time to an observed dwarf based on its location in PPS alone is not possible, especially when it is found in the inner parts of the cluster. Inspecting the histograms individually we found that most cells allow to infer multiple significant infall times. As explained in detail in Appendix A (\cref{fig:histtypes}), we can categorize each cell into one of three different classes: A - dominant recent infall population with early infall tail, B - mixed with equally significant recent and early infallen populations, and C - dominant early infallen population with recently infallen tail. We choose a threshold of \SI{3}{\giga\year} to divide \enquote{recently} and \enquote{early} infallen galaxies in each cell, as this accurately captures the separation in most cases (further motivated in the Appendix, see \cref{fig:means_of_cells}).

\begin{figure*}[ht!]
    \centerline{\includegraphics[width=\textwidth, trim={0 0 0 0}]{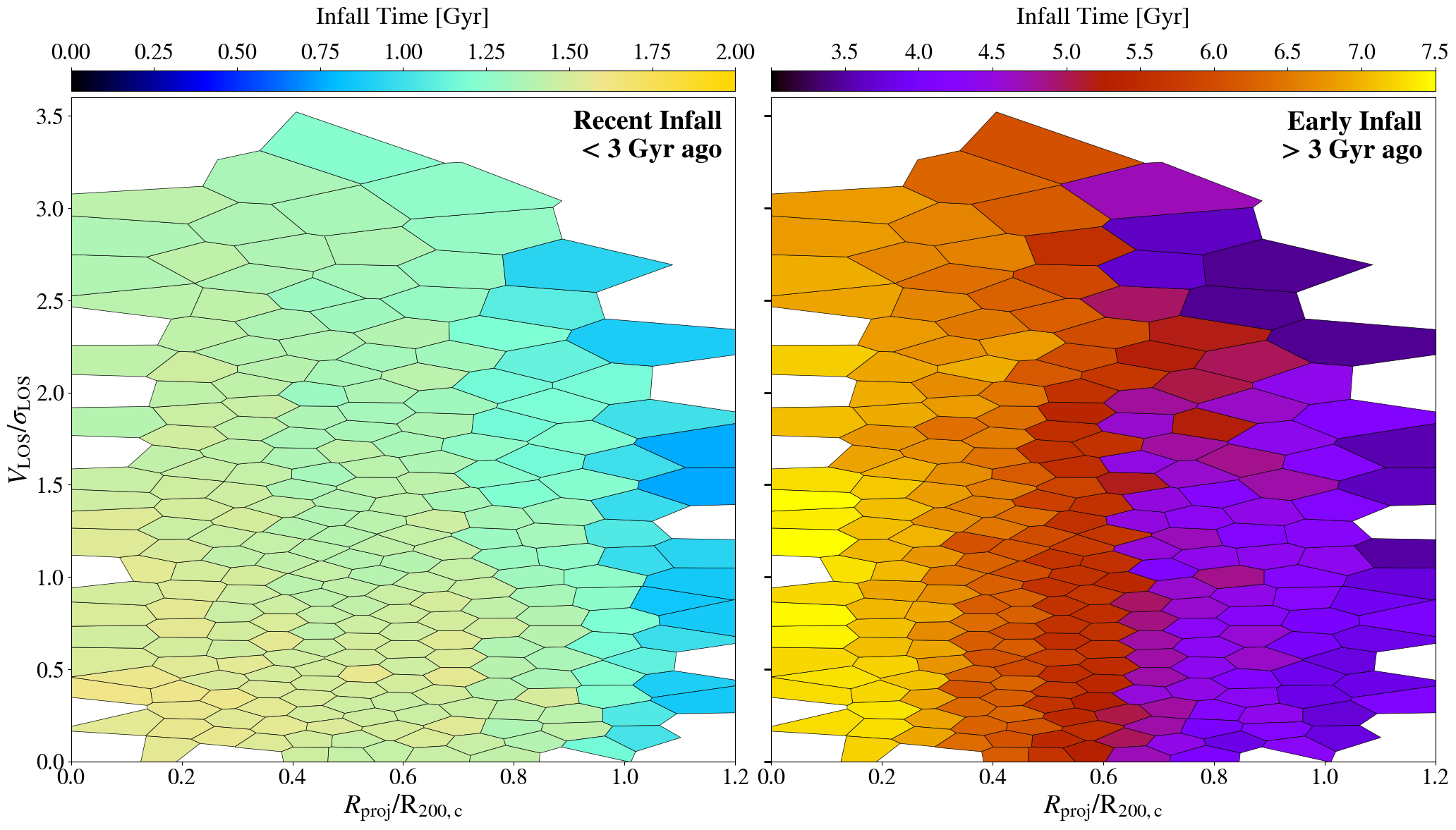}}
    \caption{Mean infall time into clusters for the dwarf galaxy population with recent vs. early accretion into cluster (left and right panel), differentiated by a threshold of \SI{3}{\giga\year}. The relative significance of the recent vs. early population in each region of PPS can be reviewed in \cref{fig:fraction_in_group}. While recently infallen galaxies are spread smoothly throughout the cluster, once these are removed there is a strong trend of the early infallen galaxies even in the cluster outskirts. The data sets, as well as programs in \textsc{Python} and \textsc{Julia} for reproducing this template, are provided in a \href{https://github.com/AnnaIvleva1/PhaseSpaceDiagnostic.git}{Github repository}.}
    \label{fig:infalltimetemplate}
\end{figure*}

\subsection{Infall time templates}\label{subsec:infalltimetemplate}

\cref{fig:infalltimetemplate} shows the PPS with the associated infall time per cell indicated by the color for recently vs. early infallen dwarf galaxies in the left and right panel, respectively. The infall time in each cell for the recent vs. early infallen galaxies is determined by the mean of the cell distribution with infall times less vs. more than \SI{3}{\giga\year}, respectively. For reference we show the determined infall times for each cell as the two dashed lines in each panel of the histogram grid in \cref{fig:histtypes}.

While there is a weak trend of increasing infall time with decreasing clustercentric distance for the recently infallen cohort (left panel), most parts of the PPS display a relatively constant mean infall time of $t_{\rm infall}\approx\SI{1.5}{\giga\year}$ throughout the whole cluster volume. The early infallen population (right panel), on the other hand, exhibits a much larger dynamic range of ages, clearly correlated with clustercentric distance (the distribution of infall times in both classes is given in the Appendix, lower panel of \cref{fig:means_of_cells}). While in the cluster center one finds galaxies with infall times of $t_{\rm infall}\geq\SI{7}{\giga\year}$, the cluster boundary at $R_{\rm proj}\sim \Rc$ hosts dwarf populations that have been accreted \SI{4}{\giga\year} ago on average in the early infall category. These are either splashback galaxies, that have returned to the cluster boundary after a pericenter passage, or objects on circular orbits. Comparing this region to the number fraction of recently infallen dwarfs shown in \cref{fig:fraction_in_group}, we find that such splashback galaxies constitute \mbox{$\sim$30\%} of the total dwarf galaxy population at $R_{\rm proj}\sim \Rc$.

Another curious note here is the comparably short time after infall ($t_{\rm infall}\sim 1.5\, \rm Gyr$) for the recent infall group at radial distances of $0.2\leq R_{\rm proj}/\Rc\leq0.5$. Comparing this region with the fraction of dwarfs belonging to the recent infall population in \cref{fig:fraction_in_group}, it follows they constitute $\sim50\%$ of all dwarfs found in this region. This ultimately demonstrates again the wide range of probable infall times possible across PPS, when comparing the same region in both (recent and early) panels.

\cref{fig:infalltimetemplate} can be used as a direct template for observed galaxies to estimate likely infall times based on the location in PPS. We demonstrated throughout this Section that providing a unique expectancy value is not possible. Instead, each region in PPS is associated with two probable infall times, being either recent ($<\SI{3}{\giga\year}$, left panel) or early ($>\SI{3}{\giga\year}$, right panel). If the observed galaxy displays signs of recent infall, e.g. active or recent star formation, large gas content or jellyfish-like stripped tails, the left panel of \cref{fig:infalltimetemplate} is applicable. On the other hand, if the galaxy lacks such features and appears red and quiescent, it is likely that it has been already exposed for significant time to the hostile cluster environment. In this case, the right panel of \cref{fig:infalltimetemplate} will provide the expected infall time for this galaxy. In addition, \cref{fig:fraction_in_group} can be used as a reference for the fraction of recently vs. early infallen galaxies that is expected in each location in PPS. In order to facilitate the applicability, we provide code snippets in \textsc{Python} and \textsc{Julia} via Github\footnote{\url{https://github.com/AnnaIvleva1/PhaseSpaceDiagnostic.git}}, which supply the data shown in \cref{fig:fraction_in_group,fig:infalltimetemplate}, as well as generate these diagrams, which observers can use as a direct comparison tool.

\section{Discussion of infall time analysis}\label{sec:discussion_infalltimes}

The statistical behavior of galaxy populations in PS and the possibility of distinct regions tied to a particular infall time has been investigated in several works. Such a trend can serve as a convenient diagnostic observers may apply to their objects, since it will hint at the extent of environmental influences a galaxy must have been exposed to in the past. This is a crucial aspect driving a galaxy's evolution, hence being an essential piece of information for explaining its present characteristics.

\subsection{Previous studies on infall time trends}\label{subsec:comparison_other_studies}

Utilizing dark matter particles as dynamical tracers for galaxies from a cosmological simulation performed by \citet{Borgani:2004}, \citet{Mahajan:2011} first identified characteristic regions in PPS that galaxies predominantly occupy depending on whether they are on their first infall, on backsplash orbits, or already virialized. Although the simulation was hydrodynamic, the authors deliberately chose to follow dark matter particles instead of the resolved baryon or dark matter halos. This approach (i) circumvented the controversy persisting across all hydrodynamic cosmological simulations, of whether baryonic subgrid models actually yield proper galaxies as found in nature and (ii) yields far better statistics because of the sheer number of available tracers. Following a similar rationale, \citet{Oman:2013} then found a superset of these regions by analyzing dark matter halos identified in a pure N-body simulation, thus being able to draw from a large cosmological volume. Employing a lower mass cut of $M_{\rm halo}\approx10^{12}\, \Msun$ at time of infall, these results could now be directly applied to observed galaxies down to stellar masses of $M_\ast\gtrsim10^{10}\,\Msun$, as the stellar-to-halo mass (halo abundance) relation is well constrained in this regime.

\citet{Rhee:2017} repeated this analysis by performing cosmological zoom-in simulations of galaxy clusters with excellent dark matter resolution, allowing them to lower the halo mass cut at infall time to $M_{\rm halo}=\eshort{3}{10}\,\Msun$. Following the same argument as before, they examined the dark matter halos instead of the baryons in order to avoid contamination of their results due to uncertainties in subgrid models. Splitting their halo sample into four cohorts with increasing infall times, they found that the maximum likelihood of these four distributions in PPS is located in different areas. This allowed them to define distinct infall time regions within the virial radius of the cluster, since before there was no clear differentiating trend reported within such \enquote{small} clustercentric distances. \citet{Pasquali:2019} further analyzed the same simulations, identifying characteristic caustic regions in PPS, which are distinguished by constant mean infall times within. Based on the halo abundance function published by \citet{Guo:2010}, \citet{Rhee:2017} stress that their results are applicable even for dwarf galaxies with stellar masses down to $M_\ast\gtrsim 10^7\,\Msun$.

While the stellar-to-halo mass relation can be well applied in the Milky Way mass regime, it is still poorly constrained below $M_{\rm halo}<10^{11}\,\Msun$, since different studies can infer vastly different slopes at these low masses \citep[][]{Wechsler:2018}. In fact, \citet{Guo:2010} predict one of the lowest stellar occupation fractions in dwarf galaxies compared to other works, while e.g. \citet{Yang:2012} lie on the opposite end, yielding a ten times more massive stellar component of $M_\ast\sim 10^8\,\Msun$ in a halo with $M_{\rm halo}=\eshort{3}{10}\,\Msun$. Additionally, all reported halo abundance functions by definition only represent the mean of the total distribution at a given halo mass. Particularly at low-mass regimes the scatter is substantial, which is one important reason for the disagreement between different studies in the first place.

\subsection{Comparison to this work}\label{subsec:newcomp}

Our approach differs in two key aspects with respect to prior works: we (i) analyzed for the first time the actual baryon halos resolved in the simulation instead of using dark matter as galaxy tracers and (ii) evaluated probable infall times of the galaxy population at a given region in PPS taking the broadness of the local distribution into account.

The former (i) methodological difference makes the result susceptible to variations in subgrid recipes, since particularly the feedback implementation will impact the baryon content of a dwarf galaxy. However, as described above, it is not straightforward to deduce the stellar component of a dark matter halo based on halo abundance matching in this small-mass regime, either. The properties of the baryon halos (galaxies) of the \textsc{Magneticum} suite have been studied extensively at this point, reproducing key scaling relations observed in the local Universe \citep[][]{Dolag:2025}. Hence we are confident that the results we report in this work represent a statistical trend for dwarf galaxies down to stellar masses of $M_\ast \sim 10^8\,\Msun$ based on our resolution-based mass cut.

The latter (ii) aspect is targeted at providing robust expectation values for the infall time of a dwarf galaxy observed at a particular location in PPS. Rather than identifying the most likely location in PPS of time-separated cohorts as done by \citet{Rhee:2017}, we specifically analyze the whole population contained within a given region. The key difference to the approach by \citet{Pasquali:2019}, on the other hand, is that we take the broadness of the infall time distribution into account when evaluating probable infall times, instead of calculating the overall mean. We demonstrated in Section \ref{subsec:spread}, that the spread in possible infall times is substantial -- particularly below $R_{\rm proj}<0.6\Rc$. In fact, this agrees well with the reported infall time histograms in the work by \citet[][c.f. their figure A1]{Pasquali:2019}, displaying very broad distributions. Their reported widths $2\sigma(\overline{T}_{\rm inf})$ of the distribution (c.f. their table 2) are very similar to our values found in \cref{fig:broadness_unimodal}, approaching large values of $\sim$\SI{5}{\giga\year}. Keeping information about the possibly multi-modal and broad infall time distribution thus poses a more optimal method for the specific task of providing a local expectancy value for the accretion time into the cluster.

As a reference, we include in the Appendix a comparison of the infall time template developed in this work to the study by \citet{Rhee:2017} in \cref{fig:infalltimes_rhee}. While we recover a similar trend in radial direction as reported by \citet{Rhee:2017}, the infall time variation across different LOS velocities turns out to be much less pronounced. This characteristic is smeared out due to mixing of multiple populations with varying ages. In principle, the longer the galaxy is orbiting in the cluster, the slower it will become due to dynamical friction, and therefore statistically exhibit lower LOS velocities. However, another galaxy can also display the same velocity at a much earlier or later time, simply by being on a different orbit. While the vertical trend disappears that way, the correlation in radial direction still persists very clearly. This is reinforced by the growth of the cluster itself -- since it accretes mass over time, also the virial radius $\Rc$ increases. This causes all objects to naturally move towards the left in the plane, since the relative projected radius $R_{\rm proj}/\Rc$ decreases.

This aspect would imply that the radial trend of infall time may be more pronounced the more massive the galaxy cluster is. So far, the extent of this issue is still unclear due to conflicting results across different studies: while \citet{Oman:2013} do report faster infall with increasing cluster mass, \citet{Rhee:2017} do not find this correlation. A similar picture holds in terms of trends with satellite mass: \citet{Rhee:2017} report similar behavior in PPS for both dwarf and giant galaxies, while the results by \citet{Oman:2013} hint at longer infall times for dwarf galaxies, which persevere longer at larger clustercentric distances due to less effective dynamical friction. Since the cosmological simulation we have analyzed in this work does not resolve large clusters, we cannot comment on trends with cluster masses. However, comparing the relative distribution in 3D PS of giant vs. dwarf galaxies (reviewed in Section \ref{sec:dmfreedwarfs}, left and middle panel of \cref{fig:spatialdistrib}), we can confirm that the two cohorts display qualitatively different distributions. An important feature is a depressed probability for dwarfs to be located at small clustercentric distances with high velocities, producing a characteristic \enquote{hump}, which perseveres also in PPS. A hint on this feature was already noticed by \citet[][c.f. their fig. 9]{Rhee:2017}, but could not be ascertained due to low number statistics. Since in our analysis we have applied a Voronoi tesselation to our distribution, we can now indeed confirm this characteristic, since each cell in \cref{fig:infalltimetemplate} by definition contains a statistically significant amount of objects.

\section{Dark matter dominated dwarf galaxies}
\label{sec:dmfreedwarfs}

Dwarf galaxies characterized by a low dark matter fraction represent a particular sub-set of the galaxy population inside clusters. Their dynamics within the cluster, as reflected by their phase-space and infall time distributions, is interlinked with possible formation scenarios like tidal striping. Additional formation mechanisms such as e.g. tidal dwarf formation might be reflected in different dynamics, which we investigate in the following.

\begin{figure}[t!]
    \centerline{\includegraphics[width=0.5\textwidth, trim={0 0 0 0}]{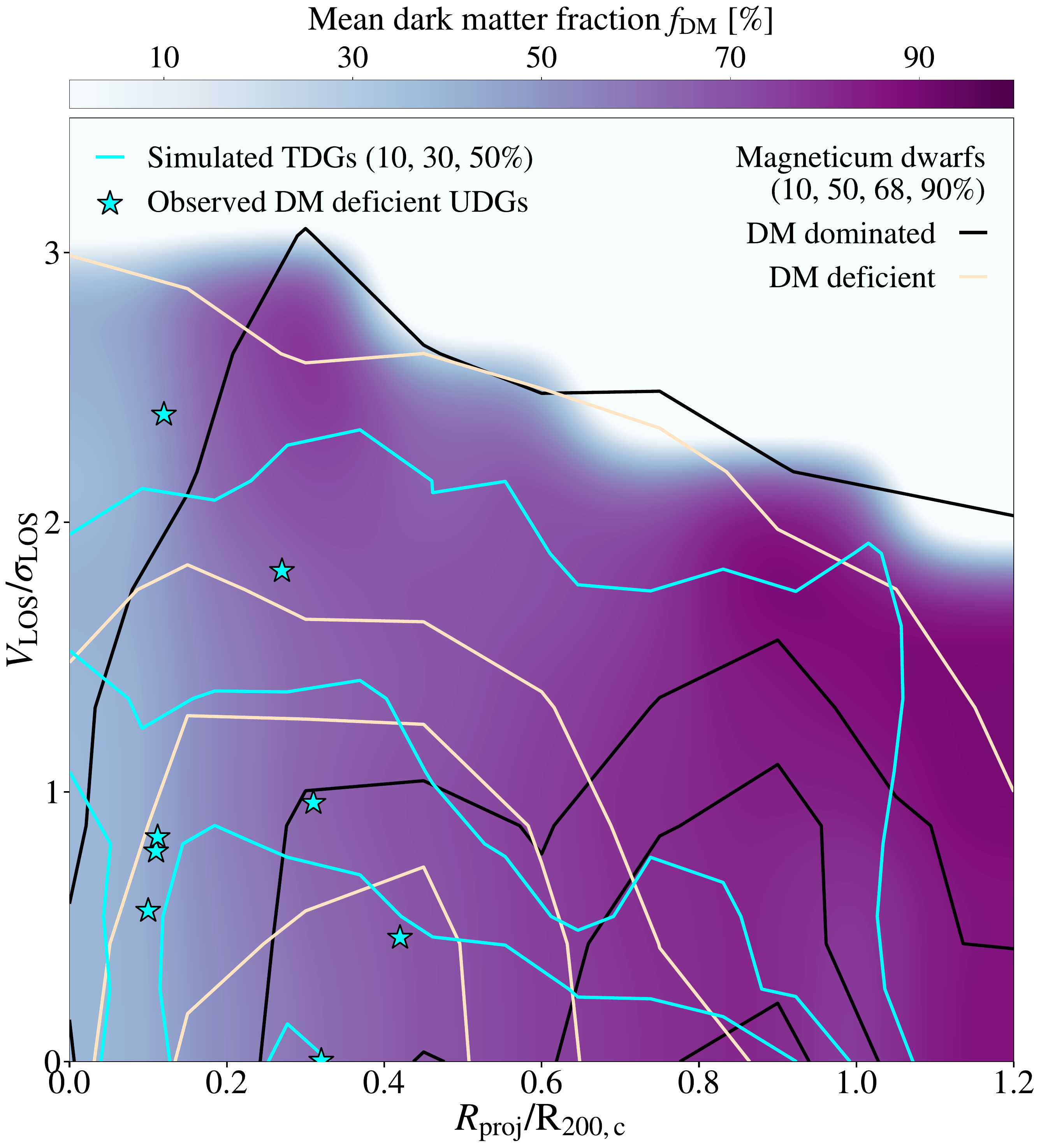}}
    \caption{Dark matter fraction of dwarf galaxies in PPS. The background shows the mean dark matter fraction of \textsc{Magneticum} dwarfs, while the black and light-orange lines show the percentiles of the dark matter-dominated and -deficient subsamples (dark matter fraction $f_{\rm DM}\gtrless90\%$). The cyan lines represent the simulated TDG sample, while the cyan stars denote dark matter-deficient UDG observations: PUDG-R15 \& -24 \citep[][]{Gannon:2022}, Yagi392 \citep[][]{Alabi:2018,Ferremateu:2018}, GMP2673 \& 2552 \citep[][]{Chilingarian:2019}, Yagi090 \& OGS1 \citep[][]{Ruiz-Lara:2018, Ferremateu:2023} and UDG32 \citep[][]{Hartke:2025}. Dwarf galaxies with low dark matter fraction driven by tidal stripping are expected to be found mostly in the inner parts of the cluster $R_{\rm proj}/\Rc\leq0.5$. If a dark-matter deficient galaxy is found at large clustercentric distances, it is likely formed via an alternative pathway, e.g. stripping of TDGs from a merger, which forms inherently dark matter-deficient galaxies in the cluster outskirts.}
    \label{fig:dmfrac}
\end{figure}

\subsection{Cosmological prediction for tidal striped dwarfs}\label{subsec:dmfrac}

The dwarf galaxy sample from {\it Box4} of the \textsc{Magneticum} simulations allows to predicted the dark matter fraction of dwarf galaxies within cosmological context. Here, \cref{fig:dmfrac} show the variations in the dark matter content of dwarf galaxies within PPS, where the background color represents the mean dark matter fraction within the pixel in PPS, as indicated by the colorbar. After being accreted into the cluster, galaxies lose dark matter (particularly from the outskirts of their extended halo) through tidal stripping while moving towards the cluster center. The outskirts of the cluster around $\Rc$ exhibit almost exclusively dark matter-dominated dwarf galaxies ($f_{\rm DM}>90\%$), with decreasing dark matter fractions at smaller radii. To highlight the relative location of the dark matter-dominated vs -deficient dwarf galaxy population in PPS, the black and light-orange contours show the $10, 50, 68 \text{ and } 90\%$ levels of objects with $f_{\rm DM}\geq90\%$ and $f_{\rm DM}<90\%$, respectively. While dark matter-dominated dwarfs are predominantly found at clustercentric distances around $R_{\rm proj}\sim 0.8\Rc$, dark matter-deficient dwarfs are most likely to be discovered around $R_{\rm proj}\approx 0.4\Rc$.

\subsection{The imprint of tidal dwarf galaxies}\label{subsec:dmfrac}

While the clustercentric trend is a well-established prediction for tidal stripping of galaxies in clusters, additional formation channels for baryon-dominated galaxies have been found by now \citep[][]{Mitrainovi:2023,Lora:2024,Ivleva:2024,Ivleva:2026}. We used the simulations presented by \citet[][see Section \ref{subsec:tdgsampleintro}]{Ivleva:2024} where it was shown that tidal dwarf galaxies (TDGs) can detach from their merger parent within a cluster environment and thus contribute to the cluster's population of dark-matter deficient objects. They can be found at significantly larger clustercentric distances, as becomes apparent from the cyan contours in \cref{fig:dmfrac}. They show $10, 30 \text{ and } 50\%$ levels of the simulated TDG sample, which were projected with the Healpix method described in Section \ref{sec:cosmo_distrib}. While overlapping with the cosmologically stripped population, they display a characteristic overdensity at intermediate to large clustercentric distances ($0.6 < R_{\rm proj}/\Rc<1$), where typically only dark-matter dominated dwarfs would be expected. Galaxy mergers are more likely to occur in the outskirts of the galaxy cluster, since they usually get accreted along a filament already in association, while the high velocity dispersion inside a cluster makes encounters between galaxies unlikely. As such, inherently dark matter-deficient TDGs are deposited in the outskirts of the cluster, subsequently moving towards the cluster center afterwards.

For comparison, we show observations of dark matter-deficient UDGs in clusters as cyan stars, where recession velocity measurements were available and the dark matter fraction is expected to be low due to a low globular cluster count \citep[][]{BurkertForbes2020}. Since these were all drawn from cluster surveys with limited coverage, they all lie within $R_{\rm proj}/\Rc<0.4$. As such, they overlay both with the cosmological dwarf and TDG sample. Considering that TDGs tend to be in the UDG regime particularly in the early phases after stripping when they are just within the virial radius \citep{Ivleva:2026}, we note here the increasing importance of probing dwarf galaxy populations at the outskirts of galaxy clusters, allowing to test simulation-based predictions.

\begin{figure*}[ht!]
    \centerline{\includegraphics[width=0.95\textwidth, trim={0 0 0 0}]{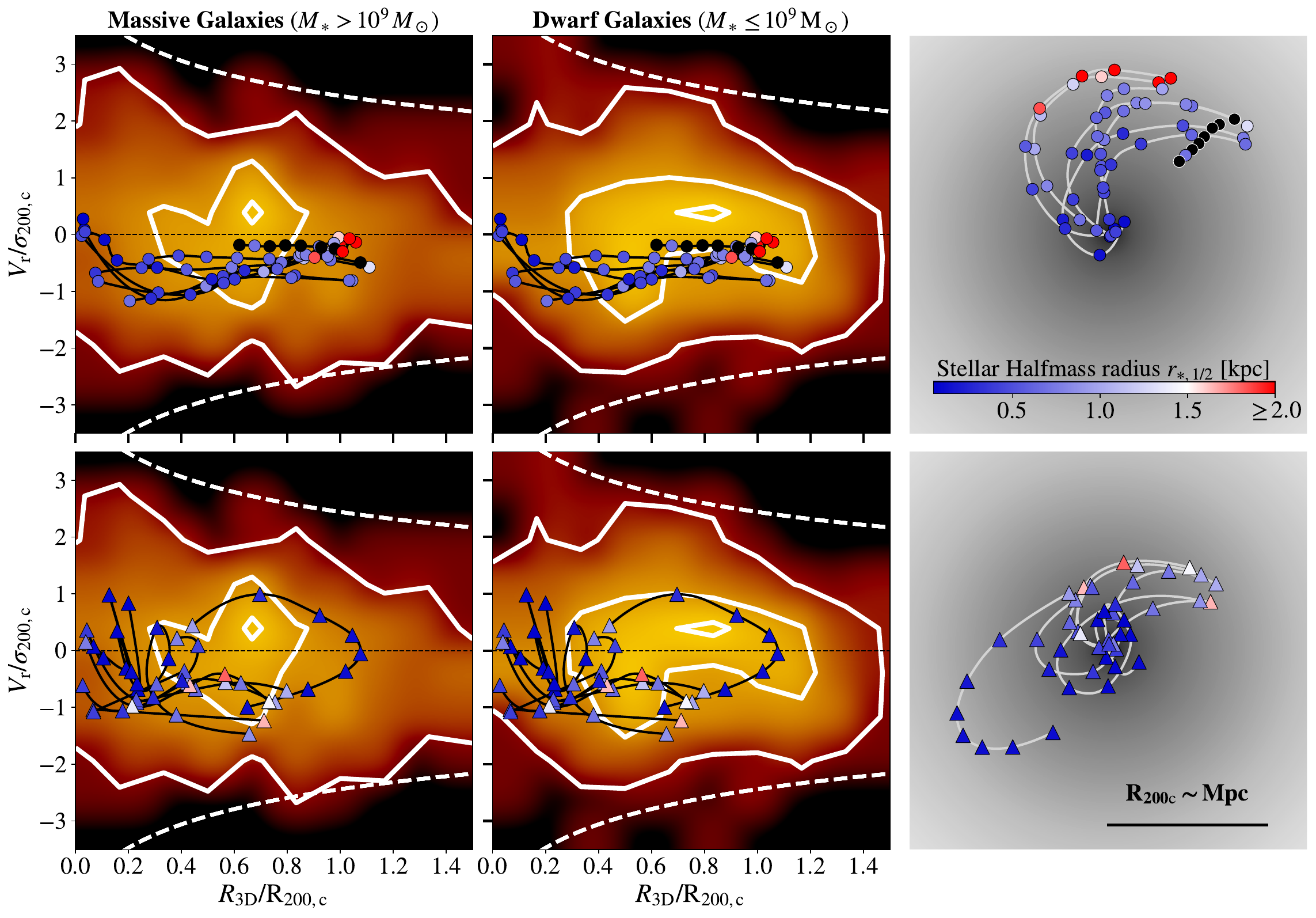}}
    \caption{3D Phase-space behavior of two simulated TDG samples with differing initial angular momentum (upper and lower row) compared to dwarf galaxies from a cosmological simulation. The background color of the left and middle column compare the distribution to giant and dwarf galaxies from \textsc{Magneticum} (10, 50 and 90\% levels are shown as white solid lines, the dashed line indicates the local escape velocity), respectively, while the right column displays the orbit undertaken by the dark matter free TDG samples. The marker color shows the dwarf's stellar halfmass radius at that moment, whereas the connecting lines indicate its time evolution over $\sim$\SI{4}{\giga\year} with timesteps of \SI{0.5}{\giga\year} (the motion is anticlockwise in the plane). Black marker color means less than six stellar particles for the given galaxy at that moment. Dwarf galaxies display a much larger spread in clustercentric distances compared to massive galaxies. The simulated dark matter-deficient galaxies acquire mostly constant radial velocities and decrease in stellar radius over time.}
    \label{fig:spatialdistrib}
\end{figure*}

\subsection{Evolution in 3D}\label{sec:bardom_3D}

The first two rows in \cref{fig:spatialdistrib} present the 3D PS, showing radial velocity $V_r$ against radial distance $R_{\rm 3D}$ to the cluster center. Similar to before, the values are normalized by the velocity dispersion $\sigma_{\rm 200,c}$ and virial radius $R_{\rm 200,c}$ of the cluster. We show the number density of cosmological samples from {\it Box4} of \textsc{Magneticum} as the orange-red background, where the left and middle column present the distribution of giant and dwarf galaxies, respectively, distinguished by a stellar mass cut ($M_\ast\lessgtr10^9\,\Msun$). The white solid lines indicate their 10, 50 and 90\% levels. The black solid lines represent for comparison the time evolution for the two simulated TDG samples by \citet[][]{Ivleva:2024} over $\sim$\SI{4}{\giga\year} with timesteps of \SI{0.5}{\giga\year} between consecutive markers, whereas the actual orbit undertaken inside the cluster in both simulations is shown in the upper and lower panel of the right column, respectively. Here, the orientation was chosen such that the depicted orbit is fully in the image plane. The marker color indicates the dwarf's current stellar halfmass radius according to the colorbar, while black means that no size can be assigned due to less than six stellar particles in the galaxy. 

For reference the white dashed lines mark the local escape velocity $v_{\rm esc} (R)$ from the cluster for a given radial \mbox{distance $R$}:
\begin{align}\label{eq:escapevel}
    v_{\rm esc} (R) &= \sqrt{\frac{2  G M_{\rm 200, c}}{ R_{200,c}} g_{c} K(R)} \quad \text{with}\\
    g_c &= \left[\ln(1+c) - c/(1+c) \right]^{-1}, \nonumber\\
    K(R) &= \frac{\ln(1+cR/\rm \Rc)}{R/\rm \Rc}, \nonumber
\end{align}
\noindent where $c, G, M_{\rm 200,c}$ and $R_{\rm 200,c}$ are the halo concentration, gravitational constant, virial mass and virial radius, respectively\footnote{\cref{eq:escapevel} follows from equipartition of the kinetic energy $v^2/2$ to the potential energy difference necessary to escape from $r=R$ to $r\rightarrow\infty$, using the NFW potential \citep[][]{NFW:1996} $\Phi(r) = -g_c K(r) G M_{\rm 200,c}/R_{\rm 200,c}$.}.

Comparing the distributions of the cosmological sample from {\it Box4} of \textsc{Magneticum}, \cref{fig:spatialdistrib} displays key differences between dwarfs and massive galaxies. First, the 90\% contours at low distances indication that massive galaxies can sustain themselves longer in the central parts with larger velocities, being more resilient against the strong tidal torques acting in the center. This was already demonstrated in detail by \citet{lotz:2019}. Second, the distribution of massive galaxies is peaked in a relatively confined clustercentric distance interval, where 50\% of them lie within $0.4\lesssim R_{\rm 3D}/\Rc \lesssim 0.8$, while dwarf galaxies have a much larger spread, where 50\% of all dwarfs occupy radial distances between \mbox{$0.2\lesssim R_{\rm 3D}/\Rc \lesssim 1.2$}. 

Instead of following regular orbits where galaxies accelerate when moving towards the cluster center, the dark matter-deficient dwarfs instead slow down and acquire almost constant radial velocities. This is driven by their gas-dominated mass reservoir in the beginning of their evolution, thus being subjected to strong ram pressure effects. The gas body of these dwarfs is therefore decelerated by the headwind, and the initially small stellar component is forced to follow the local gravitational potential defined by the gas. Due to high star formation rates relative to the dwarf main sequence, the TDGs later build a significant stellar content during the first few Gyr of their evolution, but the dwarfs have already lost most of their angular momentum by that time. As such, the distribution in PS for TDG, as well as their properties, are very different from the cosmological sample.

Over time, stars predominantly accumulate within the centers of the TDGs, because gas in the outskirts gets efficiently stripped. This is particularly the case for low effective viscosities in the ICM, leading to decreasing stellar radii over time \citep[][]{Ivleva:2026}. In the beginning, however, a large fraction of the simulated objects are very extended ($r_{\ast,1/2}>1.5\, \rm kpc$) and exhibit low surface brightnesses similar to UDGs \citep[][]{Ivleva:2024}. At that point, they are still in the outskirts of the galaxy cluster. Current low-surface brightness surveys are typically limited to the inner parts of a galaxy cluster ($R_{\rm 3D}/\Rc<0.5$). Increasing cluster coverage may allow testing the prediction emerging here, that UDGs and low-surface brightness galaxies are expected in larger numbers in the cluster outskirts around \mbox{$0.5<R_{\rm 3D}/\Rc < 1$}.

\subsection{Evolution in projection}\label{sec:bardom_proj}

\begin{figure*}[ht!]
    \centerline{\includegraphics[width=0.8\textwidth, trim={0 0 0 0}]{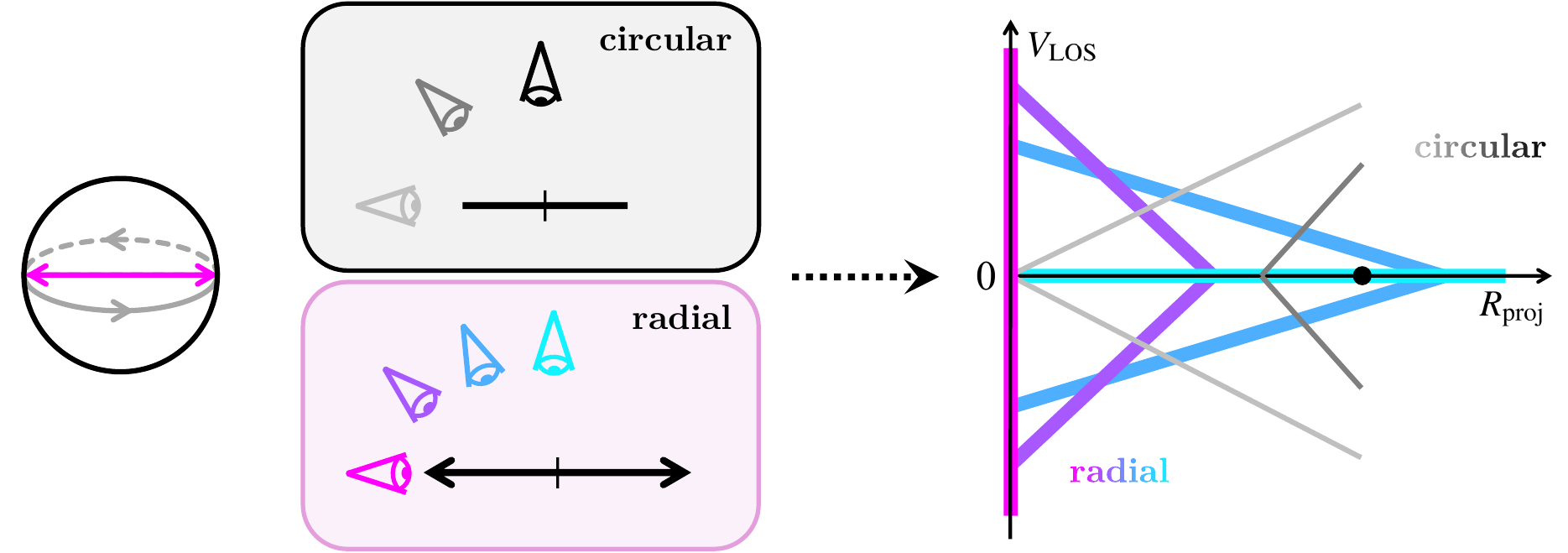}}
    \centerline{\includegraphics[width=\textwidth, trim={0 0 0 0}]{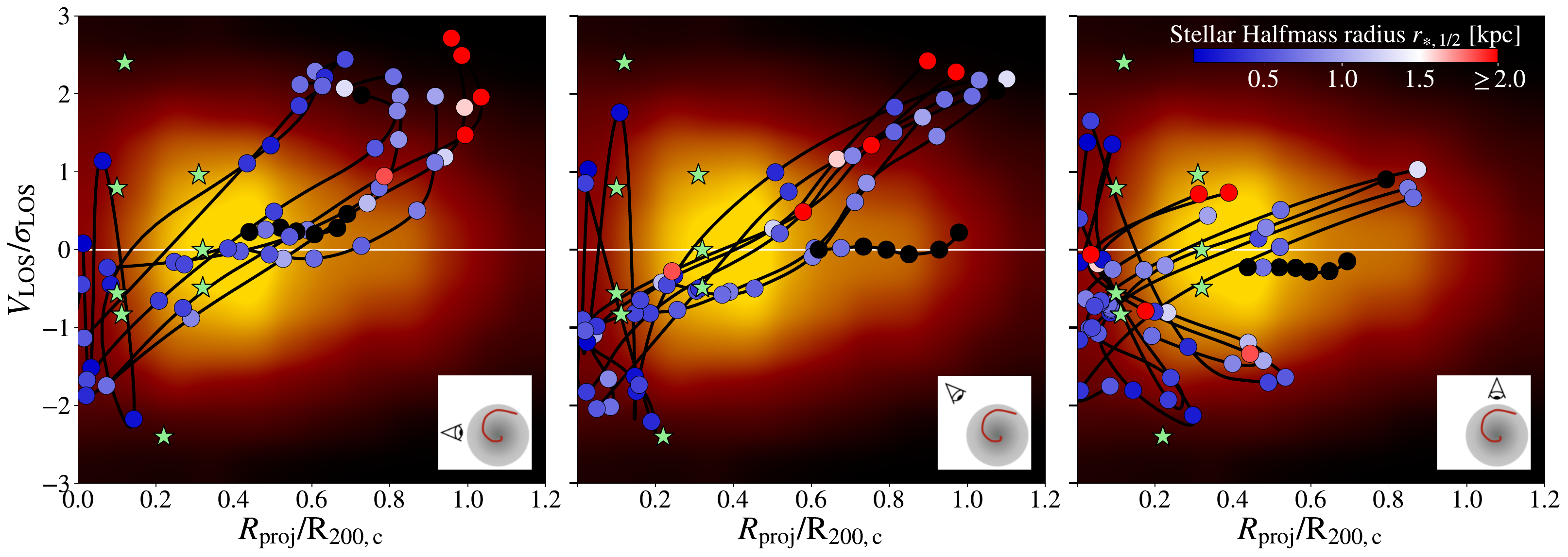}}
    \caption{\textit{Top:} Time evolution of circular and radial orbits in PPS. The line color corresponds to the observer icon color indicating the given projection. \textit{Bottom:} Time evolution over \SI{4}{\giga\year} of baryon-dominated dwarfs inside a cluster under three different projections. A dwarf's individual evolution is marked by black solid lines between the markers, where the color denotes its stellar halfmass radius and the timestep between two markers is \SI{0.5}{\giga\year}. Black marker color means less than six stellar particles for the given galaxy at that moment. The cartoon in the corner of each panel represents the projection direction. The green stars mark observations of dark matter-deficient UDGs. The background references the \textsc{Magenticum} dwarf galaxies. The simulated dark-matter deficient dwarf galaxies first move on diagonals characteristic for quasi-circular orbits, but evolve towards radial orbits in the end.}
    \label{fig:projected_showcase}
\end{figure*}

The observed location in PPS will depend not only on projection, but also on the type of orbit. In order to illustrate the possible differences, we show in the upper part of \cref{fig:projected_showcase} the projection-dependent path in PPS for two extreme orbit types: circular (gray to black) and radial (pink to cyan).

Consider a circular orbit, where the observer is positioned in the orbital plane (light gray) and the galaxy is at zero projected clustercentric distance in the beginning. It will then start to move diagonally towards the upper right in PPS as the LOS velocity component $V_{\rm LOS}$ (with respect to the inherent motion of the cluster) and the projected distance $R_{\rm proj}$ increases. When it reaches the maximum distance -- i.e., orbital radius -- the observer also measures the maximum possible LOS velocity, which is the full circular orbit velocity. Afterwards the galaxy moves back on the same path in PS, before it starts moving on the diagonal mirrored to the lower part of the diagram, as its LOS velocity becomes negative. If the observer does not lie in the orbit plane (dark gray), the diagonal paths in PPS still reach the same maximum projected distance, but naturally at smaller maximum LOS velocity. In addition, the conjuncture point between positive and negative LOS velocities moves to the right, since the galaxy does not pass the LOS connecting the observer and the cluster center anymore. Finally, the upper and lower diagonal collapse to a single point at the orbital radius with zero LOS velocity if the orbit is fully in the sky plane of the observer (black).

Turning to radial orbits, consider first an extreme case, where the motion of the galaxy is aligned with the observer's LOS (pink). The projected distance is always zero and the galaxy moves up and down the y-axis, reaching its maximum absolute LOS velocity in the midpoint of the radial orbit and crossing $V_{\rm LOS}=0$ at the two turn-around points. Now consider a slight misalignment between the orbit axis and the LOS (purple). The observer starts to see variation in the projected distance, while the maximum LOS velocity component decreases. Further tilting the observer's LOS (blue), this trend continues, until the galaxy's motion is entirely in the sky plane (cyan). At this point, the LOS velocity will always be zero, with full variation in projected distance between 0 and the maximum extent of the radial motion, i.e. the PS path is confined to the x-axis. Thus, radial orbits display a qualitatively different character in PS evolution compared to circular orbits, since they are moving on \enquote{opposite} diagonals in PS.

The TDG samples start out on initially circular and elliptical orbits respectively. However, as explained in Section \ref{sec:bardom_3D}, these objects lose angular momentum on much shorter time scales than dark matter-dominated galaxies. Therefore, they do not complete full orbital periods and instead acquire radial orbits within the first \SI{2}{\giga\year} of their evolution. This becomes apparent from the lower part of \cref{fig:projected_showcase}, where we show the evolution in PPS for the dwarfs from one of the two simulations as an example (initially on circular orbit, c.f. upper row in \cref{fig:spatialdistrib}). Each of the three panels represents a different projection, which is varied according to the cartoon in the lower right corner of each panel. The generalized orbit of the galaxies is indicated by the red line, where the gray sphere represents the cluster's ICM. As before, the marker color indicates the dwarf's stellar halfmass radius $r_{\ast,1/2}$ (black for $<6$ stellar particles), while the solid line indicates the time evolution over $\sim$\SI{4}{\giga\year} with \SI{0.5}{\giga\year} between the markers\footnote{We fitted splines to the data points for smooth evolution lines in order to make the figure more accessible. Therefore, the exact course of the lines between the markers does not necessarily represent the actual evolution, but rather is intended to guide the eye between the tracing times highlighted by the markers.}. Green stars are UDGs observed in the Perseus and Coma cluster, which could be dark matter-deficient based on their low globular cluster count (see caption of \cref{fig:dmfrac} for references). For reference, we include here the distribution of the cosmological sample from \textsc{Magneticum} as the orange-red background.

Starting out in the top right corner of PPS, the dwarfs move diagonally towards the lower left during the first few Gyr, while their orbits are still approximately circular. Their orbits become increasingly radial, until they start to move fully on the respective characteristic diagonals (purple or blue line in the upper PPS cartoon). Having no angular momentum anymore to escape the cluster center, the galaxies are then quickly destroyed by tidal torques as soon as they reach small clustercentric distances, which for most of these objects happens after about \SI{4}{\giga\year} (Appendix \ref{app:bardominfalltime}). Although the simulated galaxies are in the low-surface brightness regime \citep[$r_{\ast,1/2}\gtrsim\SI{1.5}{\kilo\parsec}$, c.f. figure 5 by][]{Ivleva:2026} relatively early in their evolution at large 3D clustercentric distances, \cref{fig:projected_showcase} demonstrates that they can easily lie in the inner regions of the cluster due to projection, coinciding with the location of observed dark matter-deficient UDGs in clusters.

\section{Summary and Conclusion}
\label{sec:sumcon}

By analyzing dwarf galaxy samples from a cosmological simulation and targeted idealized simulations of dark matter deificient galaxies, we addressed a variety of questions regarding the properties of dwarf galaxies in projected phase-space (PPS). Our findings can be summarized in three subjects: the (i) distribution of dwarfs in PPS, (ii) trends of mean infall time in PPS and (iii) differences in behavior between regular galaxies vs. dark matter-deficient halos.

(i) Inside galaxy clusters, dwarf galaxies are most likely to be observed at projected clustercentric distances of $R_{\rm proj}/\Rc\sim 0.4$ (\cref{fig:box4diagnostics}). However, the 3D distribution in radial direction is very broad, since 50\% of all dwarfs occupy radial distances between $0.2\lesssim R_{\rm 3D}/\Rc \lesssim 1.2$ (\cref{fig:spatialdistrib}). As such, dwarf galaxies populate a much larger spatial range in the cluster volume compared to Milky Way-like galaxies. This demonstrates the necessity of obtaining full coverage out to the cluster outskirts $R_{\rm 3D}/\Rc \sim 1$ for observational dwarf galaxy surveys, since a large fraction of objects is missing when focusing only on the inner parts at $R_{\rm proj}/\Rc< 0.5$.

(ii) The infall time distribution for dwarf galaxies at a given location in PPS cannot be described by a single age, because the distribution of infall times per pixel in PPS is very broad and displays variations of \mbox{$\gtrsim\SI{5}{\giga\year}$} in the inner cluster regions $R_{\rm proj}/\Rc< 0.5$ (\cref{fig:broadness_unimodal}). However, most regions are well described by a bimodal infall time distribution, which allows providing a mean infall time in each region of PPS for recently vs. early infallen galaxies hosted within. We find that using a dividing threshold of $t_{\rm infall}=\SI{3}{\giga\year}$ appropriately captures both populations. In contrast to prior works, we find equally significant recently and early infallen population at intermediate clustercentric distances \mbox{$0.2\leq R_{\rm proj}/\Rc\leq 0.5$} (\cref{fig:fraction_in_group}). In \cref{fig:infalltimetemplate} we provide a template in PPS that can be used as a direct comparison tool for estimating the probable infall time for observed samples based on the individual galaxy's location inside the diagram. The data and code allowing a comparison to observations can be recovered from a \href{https://github.com/AnnaIvleva1/PhaseSpaceDiagnostic.git}{Github repository}.

(iii) We recover a radial trend for tidal mass loss in dwarf galaxy halos, with lower dark matter fractions the closer they are to the cluster center. The most likely location to find dark matter-deficient galaxies also coincides with the maximum of the whole dwarf population at $R_{\rm proj}/\Rc\sim 0.4$ (\cref{fig:box4diagnostics}). Hence, if baryon-dominated dwarf galaxies are observed in the outer regions of cluster environments, they are unlikely to stem from tidal mass loss and could have formed via a different formation pathway. In particular, stripping of tidal dwarf galaxies occurring at the cluster boundary is an efficient mechanism for depositing such galaxies at these locations. They also display a very different behavior in PS compared to regular galaxies by acquiring almost constant radial velocities on Gyr timescales (\cref{fig:spatialdistrib}). Hence, their motion in PPS is qualitatively different from orbits of regular galaxies. The expected evolution from these latter cases -- and hence correlations between orbit type and PS location -- as well as the comparison to the behavior of baryon-dominated galaxies was illustrated in \cref{fig:projected_showcase}.

We have analyzed for the first time the dynamical behavior in PPS based on the resolved baryon halos from a cosmological simulation. Our targeted focus on the dwarf galaxy population allowed the presented analysis only for the smallest cosmological volume from the \textsc{Magneticum} suite, since the limited box size permits the required high gas resolution. However, indications for different infall time trends in PPS with varying satellite and cluster masses are accumulating. Hence, this aspect needs to be investigated through next-generation, high-resolution simulations that are large enough for assembling massive clusters while still resolving dwarf galaxies, which will be addressed in future work.

\begin{acknowledgments}
AI and KD acknowledge support by the DFG project Nr. 516355818. AI acknowledges support by the COMPLEX project from the European Research Council (ERC) under the European Union’s Horizon 2020 research and innovation program grant agreement ERC-2019-AdG 882679. LMV acknowledges support by the German Academic Scholarship Foundation (Studienstiftung des deutschen Volkes) and the Marianne-Plehn-Program of the Elite Network of Bavaria. DF thanks the ARC for financial support via DP250101673. AI, RSR, LMV, JSG and DAF acknowledge support from the German exchange program DAAD-PPP under the Project Number 57750566.
The calculations for the hydrodynamical simulations were carried out at the Leibniz Supercomputer Center (LRZ) under the project pr83li (Magneticum).
This research was supported by the Excellence Cluster ORIGINS, funded by the Deutsche Forschungsgemeinschaft under Germany's Excellence Strategy -- EXC-2094-390783311.
\end{acknowledgments}

\vspace{5mm}
\software{\textsc{Julia} \citep{Bezanson:17}, \textsc{Matplotlib} \citep{Hunter:2007}, \mbox{\textsc{HEALPix}.jl} \citep{Tomasi:2021}}

\appendix

\section{Distribution type algorithm}
\label{app:histtype_algorithm}

\begin{figure*}[ht!]
    \centerline{\includegraphics[width=0.86\textwidth, trim={0 0 0 0}]{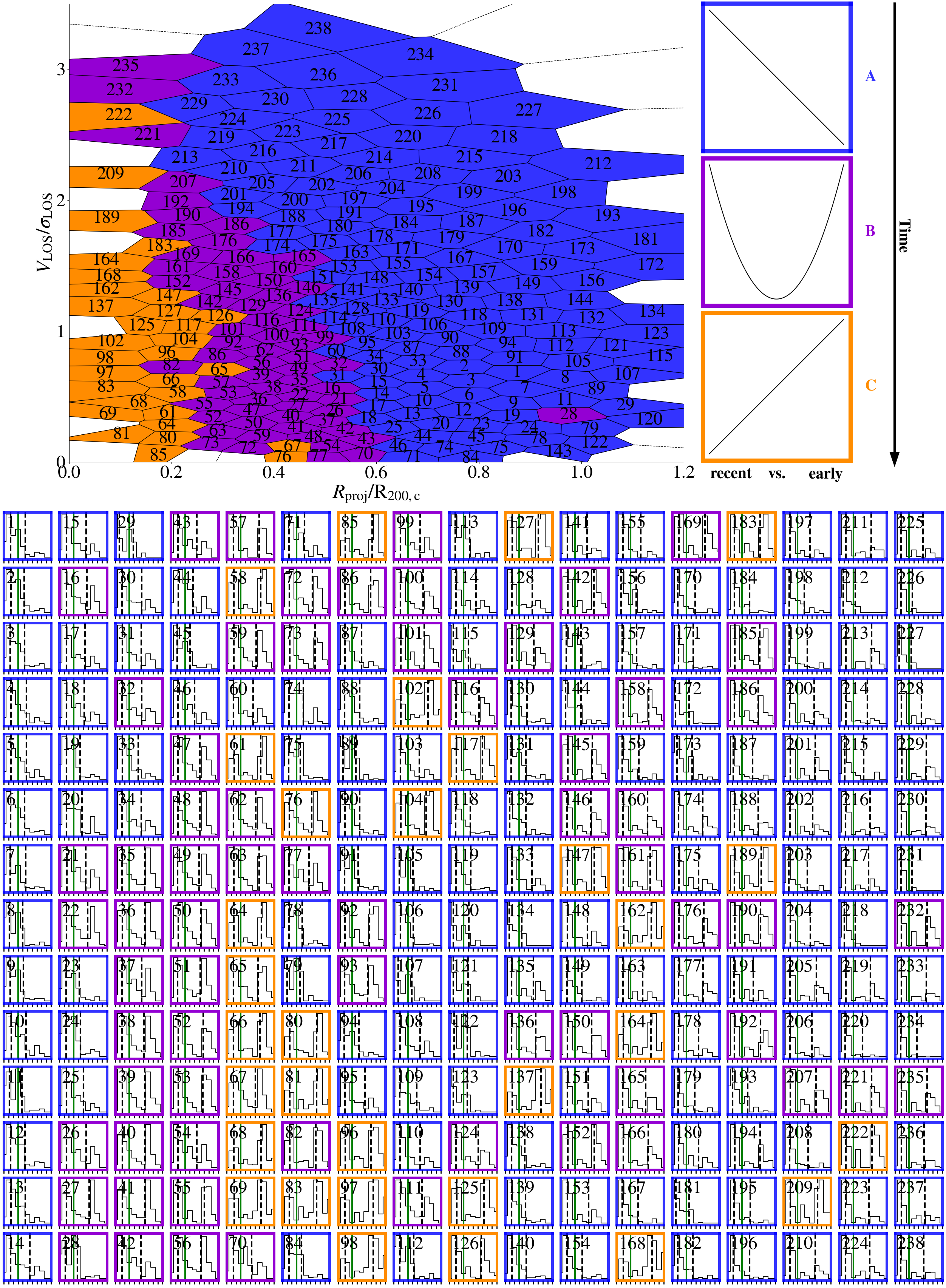}}
    \caption{Local distribution of infall times for each region in PPS. The lower grid contains the histogram for each gridcell denoted by the cellnumber shown in the upper PPS panel. Each cell is classified to belong to one of three classes (cartoon in the upper right), which differ in relative significance of recent vs. early infall, using a threshold of \SI{3}{\giga\year} (green line in histogram panels). The mean of the recent vs. early infallen population is indicated by the dashed lines in each panel. Almost the entirety of PPS displays multiple significant components in the infall time distributions. Only the very inner (outer) regions are dominated by mostly early (recently) infallen galaxies.}
    \label{fig:histtypes}
\end{figure*}

The upper panel of \cref{fig:histtypes} shows the constructed grid of dwarf galaxies within PPS, where each cell contains \mbox{$\sim$1000} galaxies. Additionally, the distribution of infall times within each cell is plotted below, where the number in the upper left corner of each panel indicates the respective cell from PPS above. The ticks in each histogram are $0-\SI{10}{\giga\year}$ with steps of \SI{1}{\giga\year}. Inspecting the histograms individually, it follows that most cells contain mixed populations with multiple significant infall times. This is particularly extreme for the innermost parts of the cluster, explaining the large spread of infall times found in \cref{fig:broadness_unimodal}.

In principle, more than two populations can contribute to the final distribution per cell. However, examining the panels yields that most cases can be described by one or two modes. As such we can limit our classification scheme to a mixture of two populations. We categorize each cell into one of three different classes, as indicated by the three vertical panels in the upper right of \cref{fig:histtypes}: A - dominant recent infall population with early infall tail (blue), B - mixed with equally significant recent and early infallen populations (purple), and C - dominant early infallen population with recently infallen tail (orange).

In order to classify each cell in PPS, we apply elementary checks regarding the shape of the infall time distribution. This is achieved by dividing the histogram of infall times into three regions (left, right and around the mean $t_{\rm mean}$) and count how many galaxies $N$ lie in each region compared to the total number $N_{\rm tot}$ of galaxies in the cell. In order to avoid contamination by extreme outliers here, we evaluate the mean $t_{\rm mean}$ after excluding all galaxies with infall times that are smaller and larger than the 5th and 95th percentiles ($t_{5}$ and $t_{95}$), respectively. Thus we arrive at three numbers:

\begin{align*}
    N_{\rm center} &= N(t_{\rm mean} - \Delta t/2 \leq t \leq t_{\rm mean} + \Delta t/2)\\
    N_{\rm left} &= N(t\leq t_{\rm mean})\\
    N_{\rm right} &= N(t_{\rm mean} \leq t)
\end{align*}

\noindent where $\Delta t = 0.4 \,(t_{95} - t_{5})$ is taken as a proxy for the \enquote{central} region of the distribution. The condition for each of the classes A, B or C (cartoon in upper right of \cref{fig:histtypes}) are

\begin{align*}
    N_{\rm left}&\geq 0.57 \,N_{\rm tot} \longrightarrow A\\
     N_{\rm center}&< 0.35 \,N_{\rm tot} \longrightarrow B  \\
     N_{\rm right}&\geq 0.57 \,N_{\rm tot} \longrightarrow C
\end{align*}

The weighting factors for each category were chosen such that the algorithm can assign a class to as many cells as possible, without being overly lax, such that one would reach the same conclusion when classifying by visual inspection.

\begin{figure}[t!]
    \centerline{\includegraphics[width=0.45\textwidth, trim={0 0 0 0}]{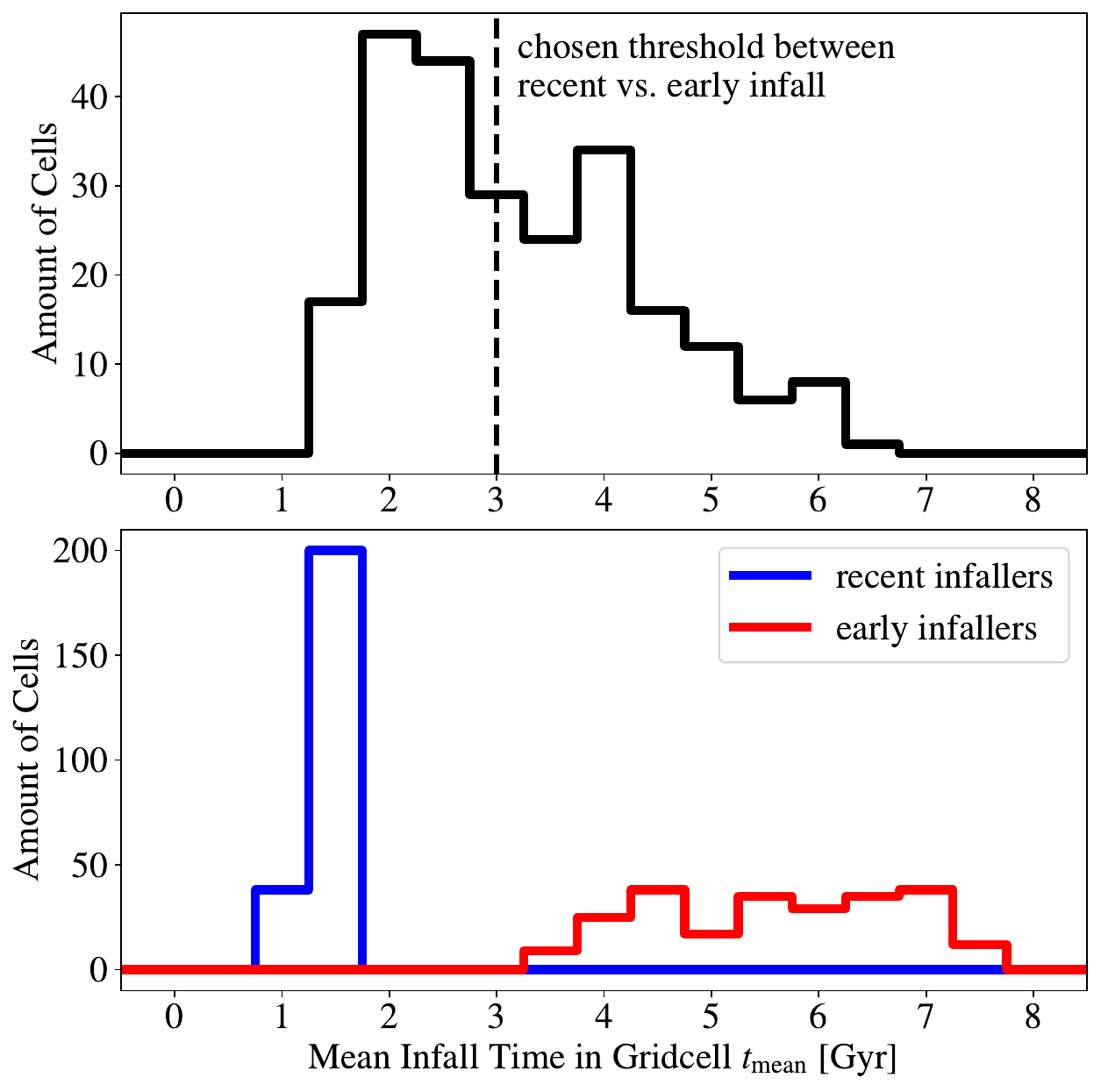}}
    \caption{\textit{Top:} Histogram of mean infall times $t_{\rm mean}$ in each cell in PPS, when assuming simple mean over full distribution. \textit{Bottom:} Distribution of infall times in recently vs. early infallen galaxies of each grid cell.}
    \label{fig:means_of_cells}
\end{figure}

In principle, the algorithm is based on the shape of the distribution around a dynamic threshold $t_{\rm mean}$, which can vary between different cells. In order to provide an applicable infall time template, however, it is more straightforward to use a fixed threshold dividing the recent and early infallen galaxies. Inspecting the upper histogram of $t_{\rm mean}$ from each Voronoi cell in \cref{fig:means_of_cells}, it follows that $t_{\rm infall}=3\,\rm Gyr$ represents an appropriate division. The lower histogram finally shows the inferred distribution of infall times in recently vs. early infallen galaxies in each gridcell, using this threshold.

This classification reveals three distinct regions in PPS that are associated to the three classes. At $R_{\rm proj}/\Rc\gtrsim 0.6$, most dwarfs have fallen in recently ($t_{\rm infall}<\SI{3}{\giga\year}$), though a tail of populations with early infall  ($t_{\rm infall}>\SI{3}{\giga\year}$) can already be observed (A). This switches to a well mixed region between $0.2\lesssim R_{\rm proj}/\Rc\lesssim 0.6$, where recently and early infallen dwarf galaxies are equally probable to be found (B). Finally, the innermost regions of PPS are dominated by the oldest population, accreted by the cluster early on (C). Interestingly, this area with most dominant old dwarfs is confined to the innermost cluster regions with $R_{\rm proj}/\Rc\lesssim 0.2$. In contrast, a quite large volume until intermediate clustercentric distances where many spectroscopic dwarf galaxy surveys are currently focused on ($R_{\rm proj}/\Rc\lesssim 0.5$), hosts a significant number of recently infallen dwarf galaxies with $t_{\rm infall}<\SI{3}{\giga\year}$. This is driven both by galaxies with radial orbits, causing them to reach the cluster center on short timescales, as well as interlopers appearing in the inner parts of the cluster due to projection.

\section{Infall time template comparison}
\label{app:histtype_algorithm}
\begin{figure*}[ht!]
    \centerline{\includegraphics[width=.925\textwidth, trim={0 0 0 0}]{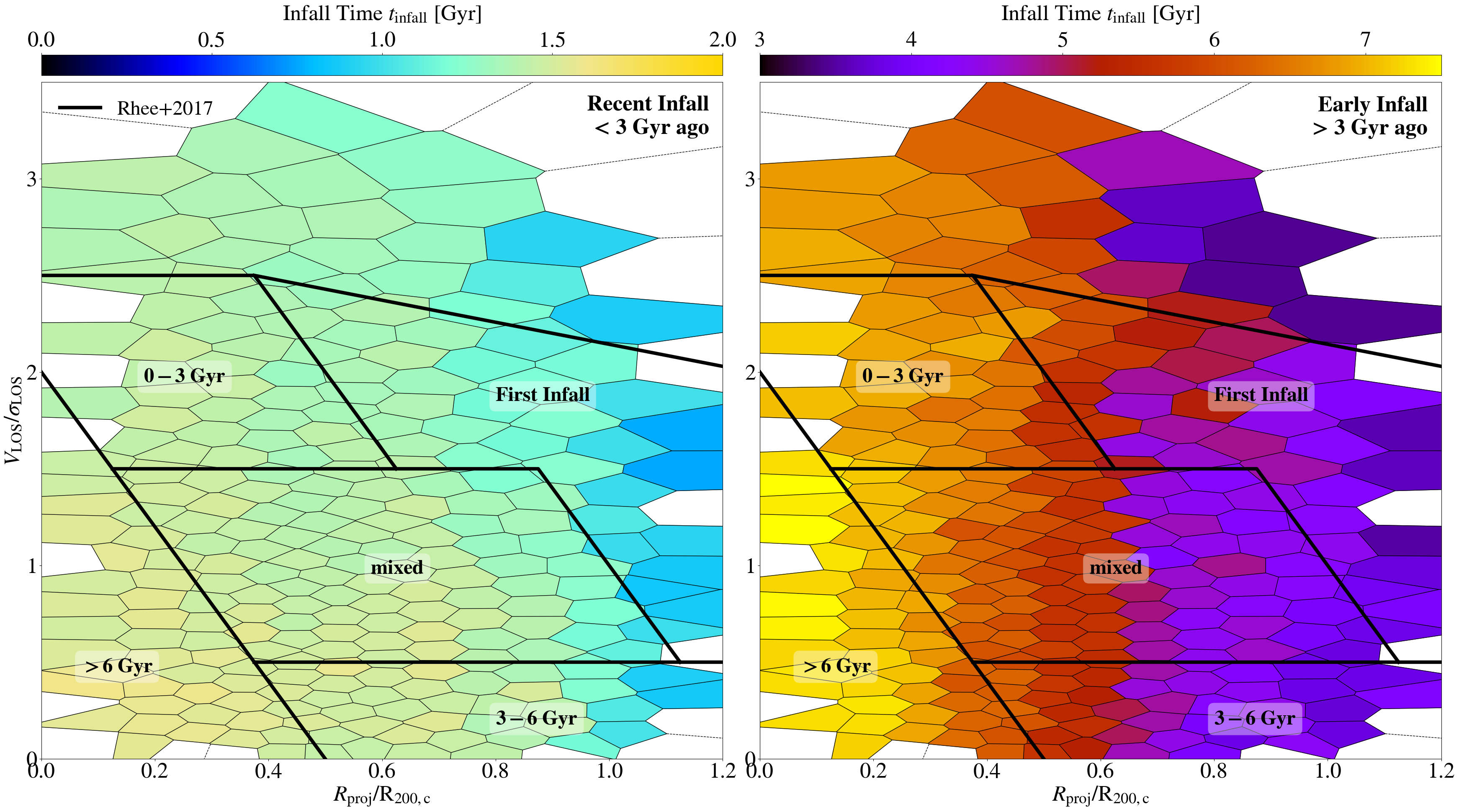}}
    \caption{Identical to the infall time template shown in \cref{fig:infalltimetemplate}, with additional comparison to the result by \citet{Rhee:2017}, who classified regions in PPS according to expected infall time (black lines and annotations).}
    \label{fig:infalltimes_rhee}
\end{figure*}

\cref{fig:infalltimes_rhee} presents a comparison between the infall time template found in our work compared to the results by \citet{Rhee:2017}. We notice that there is a significant amount of dwarf galaxies inside at high velocities ($V_{\rm LOS}/\sigma_{\rm LOS}>2.5$), which are in excess when comparing to the upper edge of the infall regions by \citet{Rhee:2017}. We are unsure what causes this discrepancy, since there is no apparent reason for these regions to be less populated when changing cluster or galaxy masses. However, we note that such a cut-off in vertical direction might be caused when selecting only galaxies with LOS velocities lower than the local radial escape velocity $v_{\rm esc} (R)$ at radial distance $R$ (\cref{eq:escapevel}). Since all galaxies in our sample are confirmed to be bound to the galaxy cluster by \textsc{Subfind} (Section \ref{subsec:magneticum}), we did not apply such a selection criterion.

\section{Infall time diagnostics for baryon-dominated dwarfs}\label{app:bardominfalltime}

\begin{figure*}[ht!]
    \centerline{\includegraphics[width=.92\textwidth, trim={0 0 0 0}]{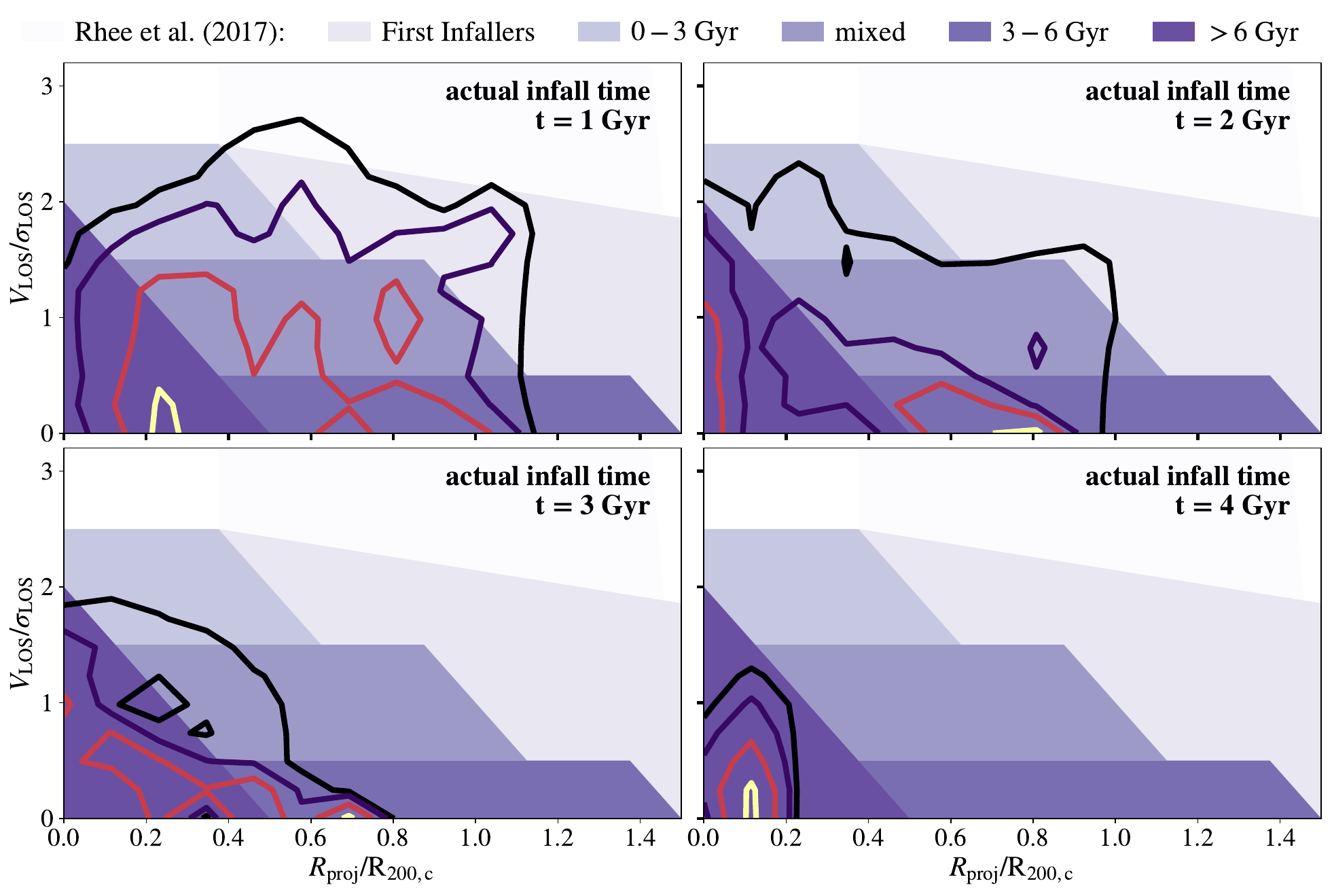}}
    \caption{Expected distribution in PPS of baryon-dominated dwarfs. The four panels show the TDG sample at four different times after infall into the cluster as noted in the corner, where the contours mark the 5, 20, 50, 90-percentiles of the distribution after applying $\sim$800 projections. The purple background represents the infall time regions defined by \citet{Rhee:2017}.}
    \label{fig:evol_contours}
\end{figure*}

\cref{fig:evol_contours} presents a sequence of the momentary distribution in PPS of the TDG samples (Section \ref{subsec:tdgsampleintro}) at four times, being 1,2,3 and \SI{4}{\giga\year} after infall into the cluster as noted in the upper right corner of each respective panel. The contours represent 5, 20, 50, 90-percentiles of the distribution, which stems from the same projection method as applied in Section \ref{sec:cosmo_distrib} ($\sim$800 projections with equal probability). For comparison we show the infall time diagnostic by \citet{Rhee:2017} as the purple background. Early-on, the distribution has a large spread both in distance and velocity, while the dwarf's orbits are still mostly circular. This turns into a distribution which is dominated by radial orbits by $t=\SI{3}{\giga\year}$, covering the characteristic lower left region of the plane (top of \cref{fig:projected_showcase}). A Gyr later, all dwarfs have already reached the central region of the cluster, thus being observed exclusively in the lower left corner under any projection.

Due to their fast depletion of angular momentum driven by ram pressure, all such baryon-dominated dwarfs reach the cluster center much quicker than their dark matter-dominated counterparts. Similar to the young population of cosmological dwarf galaxies defined in \cref{fig:infalltimetemplate}, these dwarfs thus do not follow the standard infall time regions defined by \citet{Rhee:2017}, being much younger than predicted based on their PPS location. 

\bibliography{bib}
\bibliographystyle{apj_url}

\end{document}